\documentclass[longauth]{aa}
\usepackage{graphicx}
\usepackage{txfonts}
\usepackage{natbib} 
\usepackage{url} 
\usepackage{multirow} 
\def\hr{\hbox{$^{\rm h}$}}                 
\def\deg{\hbox{$^\circ$}}
\def\fdeg{\hbox{$.\!\!^\circ$}}            
\def\fsec{\hbox{$.\!\!^{\rm s}$}}          
\def\arcm{\hbox{$^\prime$}}                
\def\arcs{\hbox{$^{\prime\prime}$}}        
\def\farcs{\hbox{$.\!\!^{\prime\prime}$}}  
\def\mm{\hbox{$^{\rm m}$}}                 
\def \agile {\textit{AGILE}}
\def \ergsec{\hbox{erg\,s$^{-1}$}}

\def \phcmsec{\hbox{photons\,cm$^{-2}$\,s$^{-1}$}}

\def \gray {$\gamma$-ray}
\def \source {\hbox{GB\,1310$+$487}}

\begin{document}


   \title{Two active states of the narrow-line gamma-ray-loud AGN GB\,1310$+$487}


   \author{K.~V.~Sokolovsky\inst{1,2,3}\fnmsep\thanks{e-mail:~\texttt{kirx@scan.sai.msu.ru}}
          \and
          F.~K.~Schinzel\inst{1,4}
          \and
          Y.~T.~Tanaka\inst{5}%
          \and
          P.~K.~Abolmasov\inst{3}
          \and
          E.~Angelakis\inst{1}
          \and
          A.~Bulgarelli\inst{6}
          \and
          L.~Carrasco\inst{7}
          \and
          S.~B.~Cenko\inst{8,9} 
          \and
          C.~C.~Cheung\inst{10}
          \and
          K.~I.~Clubb\inst{8} 
          \and
          F.~D'Ammando\inst{11,12,13} %
          \and
          L.~Escande\inst{14} %
          \and
          S.~J.~Fegan\inst{15}
          \and
          A.~V.~Filippenko\inst{8} 
          \and
          J.~D.~Finke\inst{16}
          \and
          L.~Fuhrmann\inst{1}
          \and
          Y.~Fukazawa\inst{17}%
          \and
          E.~Hays\inst{9}%
          \and
          S.~E.~Healey\inst{18}%
          \and
          Y.~Ikejiri\inst{17} 
          \and
          R.~Itoh\inst{17}%
          \and
          K.~S.~Kawabata\inst{5} 
          \and
          T.~Komatsu\inst{17} 
          \and
          Yu.~A.~Kovalev\inst{2}
          \and
          Y.~Y.~Kovalev\inst{2,1}
          \and
          T.~P.~Krichbaum\inst{1}
          \and
          S.~Larsson\inst{19,20,21}
          \and
          M.~L.~Lister\inst{22}
          \and
          B.~Lott\inst{23,24}
          \and
          W.~Max-Moerbeck\inst{25} 
          \and
          I.~Nestoras\inst{1}
          \and
          C.~Pittori\inst{26}
          \and
          T.~Pursimo\inst{27}
          \and
          A.~B.~Pushkarev\inst{28,29,1}
          \and
          A.~C.~S.~Readhead\inst{25}%
          \and
          E.~Recillas\inst{7}
          \and
          J.~L.~Richards\inst{22}%
          \and
          D.~Riquelme\inst{30}
          \and
          R.~W.~Romani\inst{18}%
          \and
          K.~Sakimoto\inst{17} 
          \and
          M.~Sasada\inst{31} 
          \and
          R.~Schmidt\inst{1}
          \and
          M.~S.~Shaw\inst{18}
          \and
          A.~Sievers\inst{30}
          \and
          D.~J.~Thompson\inst{9}
          \and
          M.~Uemura\inst{5} 
          \and
          H.~Ungerechts\inst{30}
          \and
          S.~Vercellone\inst{32} 
          \and
          F.~Verrecchia\inst{26}
          \and
          M.~Yamanaka\inst{33} 
          \and
          M.~Yoshida\inst{5} 
          \and
          J.~A.~Zensus\inst{1}
          }

   \institute{
Max-Planck-Institut f\"ur Radioastronomie, Auf dem H\"ugel 69, D-53121 Bonn, Germany
\and
Astro Space Center of Lebedev Physical Institute, Profsoyuznaya Str. 84/32, 117997 Moscow, Russia
\and
Sternberg Astronomical Institute, Moscow State University, Universitetskii~pr. 13, 119992 Moscow, Russia
\and
Department of Physics and Astronomy, University of New Mexico, Albuquerque NM, 87131, USA
\and
Hiroshima Astrophysical Science Center, Hiroshima University, 1-3-1, Kagamiyama, Higashi-Hiroshima, Hiroshima 739-8526, Japan
\and
INAF/IASF--Bologna, Via Gobetti 101, I-40129 Bologna, Italy
\and
Instituto Nacional de Astrofisica, Optica y
Electronica, Tonantzintla, Puebla, C.P. 72860 Mexico
\and
Department of Astronomy, University of California, Berkeley, CA 94720-3411, USA
\and
NASA/Goddard Space Flight Center, Greenbelt, MD 20771, USA
\and
National Research Council Research Associate, National
Academy of Sciences, Washington, DC 20001, resident at Naval Research
Laboratory, Washington, DC 20375, USA
\and
Dip. di Fisica, Universit\'a degli Studi di Perugia, Via A. Pascoli,
I-060123 Perugia, Italy
\and
INFN -- Sezione di Perugia, Via A. Pascoli, I-06123 Perugia, Italy
\and
INAF-IRA Bologna, Via Gobetti 101, Bologna, Italy
\and
Universit{\'e} Bordeaux 1, CNRS/IN2p3, Centre d'{\'E}tudes Nucl{\'e}aires de Bordeaux
Gradignan, 33175 Gradignan, France
\and
Laboratoire Leprince-Ringuet, \'Ecole polytechnique, CNRS/IN2P3, Palaiseau, France
\and
U.S.\ Naval Research Laboratory, Code 7653, 4555 Overlook Ave. SW,
Washington, DC, 20375-5352, USA
\and
Department of Physical Sciences, Hiroshima University, Higashi-
Hiroshima, Hiroshima 739-8526, Japan
\and
Department of Physics, Stanford University, Stanford, CA 94305, USA
\and
Department of Astronomy, Stockholm University, SE-106 91 Stockholm, Sweden
\and
The Oskar Klein Centre for Cosmoparticle Physics, AlbaNova, SE-106 91 Stockholm, Sweden
\and
Department of Physics, Stockholm University, AlbaNova, SE-106 91 Stockholm, Sweden
\and
Department of Physics, Purdue University, 525 Northwestern Ave, West Lafayette, IN 47907, USA
\and
Univ. Bordeaux, CENBG, UMR 5797, F-33170 Gradignan, France
\and
CNRS, IN2P3, CENBG, UMR 5797, F-33170 Gradignan, France
\and
Cahill Center for Astronomy and Astrophysics,
California Institute of Technology,
1200 E. California Blvd.,
Pasadena, CA 91101,
USA
\and
ASI--ASDC, Via G. Galilei, I-00044 Frascati (Roma), Italy  
\and
Nordic Optical Telescope, Apdo. de Correos 474, 38700 Santa Cruz de la Palma, Spain
\and
Pulkovo Observatory, Pulkovskoe Chaussee 65/1, 196140 St.~Petersburg, Russia
\and
Crimean Astrophysical Observatory, 98409 Nauchny, Crimea, Ukraine
\and
Instituto Radioastronom\'{i}a Milim\'{e}trica,, Avenida Divina Pastora 7,         
Local 20, E 18012, Granada, Spain
\and
Department of Astronomy, Kyoto University, Kitashirakawa-Oiwake-cho, Sakyo-ku, Kyoto 606-8502, Japan
\and
INAF/IASF--Palermo, Via U.~La Malfa 153, I-90146 Palermo, Italy
\and
Kwasan Observatory, Kyoto University, Ohmine-cho Kita Kazan, Yamashina-ku, Kyoto, 607-8471, Japan
           }

   \date{Received November 8, 2012; accepted January 8, 2014}

 
  \abstract
   {Previously unremarkable, the extragalactic radio source 
GB\,1310$+$487 showed a $\gamma$-ray flare on 2009 November~18, reaching a
daily flux of $\sim 10^{-6}$\,photons\,cm$^{-2}$\,s$^{-1}$ at energies $E>100$\,MeV and 
became one of the brightest GeV sources for about two weeks. 
Its optical spectrum shows strong forbidden-line
emission while lacking broad permitted lines, which is not typical for a
blazar. Instead, the spectrum resembles those of narrow emission-line galaxies.}
   {We investigate changes in the object's radio-to-GeV spectral energy distribution (SED) 
during and after the prominent $\gamma$-ray flare with the 
aim of determining the nature of the object and of constraining
the origin of the variable high-energy emission.}
   {The data collected by the Fermi and AGILE satellites at $\gamma$-ray
energies;
Swift at~X-ray and ultraviolet (UV); the Kanata, NOT, and
Keck telescopes at optical; OAGH and WISE at infrared (IR); and IRAM~30\,m, OVRO~40\,m, 
Effelsberg~100\,m, RATAN-600, and VLBA at radio are analyzed together to trace
the SED evolution on timescales of months.}
   {The $\gamma$-ray/radio-loud narrow-line active galactic nucleus (AGN) is located at redshift $z=0.638$.
It shines through an unrelated foreground galaxy at $z=0.500$. The AGN
light is probably amplified 
by gravitational lensing. The AGN SED shows a two-humped structure typical of blazars
and $\gamma$-ray-loud narrow-line Seyfert~1 galaxies, with
the high-energy (inverse-Compton) emission dominating by more than an order of magnitude 
over the low-energy (synchrotron) emission during $\gamma$-ray flares. The
difference between the two SED humps is smaller during the low-activity state. 
Fermi observations reveal a strong correlation between the
$\gamma$-ray flux and spectral index, with the hardest spectrum observed
during the brightest $\gamma$-ray state. 
The $\gamma$-ray flares occurred before and during a slow rising trend in the
radio, but no direct association between $\gamma$-ray and radio
flares could be established.}
   {If the $\gamma$-ray flux is a mixture of synchrotron self-Compton (SSC) and
external Compton (EC) emission, the observed GeV spectral variability may
result from varying relative contributions of these two emission
components. This explanation fits the observed changes in the overall IR to $\gamma$-ray SED.}


   \maketitle
%

\section{Introduction}
\label{intro}

Blazars are active galactic nuclei (AGNs) in which relativistically beamed emission from the 
jet dominates the radiative output across most of the electromagnetic spectrum.
The spectral energy distribution (SED) of a blazar has two broad components: 
one peaking between far-IR and X-ray wavelengths and the other peaking at 
$\gamma$-rays \citep{2010ApJ...716...30A}.
The low-energy emission component is believed to be dominated 
by synchrotron radiation of relativistic electrons/positrons  
in the jet. Radiation at higher energies could be due to the inverse-Compton
scattering of synchrotron photons emitted by the electrons themselves
(synchrotron self-Compton process, SSC; e.g., 
\citealt{1974ApJ...188..353J,1989ApJ...340..181G,1996A&AS..120C.537M}) and/or photons from 
external sources (external Compton process, EC; e.g., 
\citealt{1994ApJ...421..153S,2002ApJ...575..667D}). The sources of the 
external seed photons for the EC process include the accretion disk, 
broad-line region (BLR) clouds, warm dust (dusty torus), synchrotron
emission from other faster/slower regions of the jet, and the cosmic microwave
background (CMB), with their relative contributions varying for different objects.
The models based on inverse-Compton scattering by relativistic electrons
are generally referred to as leptonic models
\citep{2008MNRAS.385..283C,2009MNRAS.397..985G,2010arXiv1006.5048B,2012arXiv1205.0539B}.
An alternative view regarding the origin of blazar high-energy emission is
represented by hadronic models
\citep{2001APh....15..121M,2003APh....18..593M,2011IAUS..275...59S}, where relativistic protons 
in the jet are the primary accelerated particles. 
We adopt the leptonic models as the basis for the following discussion.

Two types of radio-loud AGNs give rise to the blazar phenomenon: 
flat-spectrum radio quasars (FSRQs) and BL~Lacertae-type objects (BL~Lacs). 
Flat-spectrum radio quasars are characterized by high luminosities, prominent broad emission lines in their optical spectra,
and the peak of synchrotron jet emission occurring at mid- or far-IR wavelengths.  
Thermal emission, probably originating in the accretion disk surrounding
the central black hole, may contribute a significant fraction of the optical
and UV emission in some FSRQs
\citep{2006A&A...453..817V,2009MNRAS.400.1521J,2010ApJ...721.1425A}.
BL~Lacertae-type objects, on the other hand, show mostly featureless
optical spectra dominated by the nonthermal continuum produced by
a relativistic jet. Their synchrotron emission peak is located 
between far-IR and hard-X-ray energies
\citep{1995ApJ...444..567P,1998MNRAS.299..433F,1998MNRAS.301..451G}. 
In GeV $\gamma$-rays, BL~Lacs show a wide
distribution of spectral slopes, while FSRQs almost exclusively exhibit soft
$\gamma$-ray spectra \citep{2010ApJ...715..429A}.
It is not clear whether there is a physical distinction between BL~Lacs and
FSRQs, or if they represent two extremes of a continuous distribution of AGN
properties such as black hole mass ($M_\bullet$), spin, or accretion rate
\citep{2011MNRAS.tmp..627G}.
Recently, five radio-loud narrow-line Seyfert~1 galaxies (NLSy1s) have been detected in 
$\gamma$-rays by {\em Fermi}/LAT, suggesting the presence of a new class of 
$\gamma$-ray-emitting AGNs \citep{2009ApJ...707L.142A,2012MNRAS.426..317D}.
The relationship between NLSy1 and blazars is under debate. 
It has been suggested that radio-loud NLSy1 galaxies harbor
relativistic jets \citep{2013arXiv1301.5785F,2013arXiv1303.3030D}, 
but unlike blazars they are powered by less massive black holes hosted by spiral galaxies 
\citep{2008ApJ...685..801Y,2013arXiv1302.2942K}. 
The presence of a relativistic jet is supported by observation of 
superluminal motions in the parsec-scale radio jet of the
NLSy1 SBS\,0846$+$513 \citep{2012MNRAS.426..317D}. The observational
evidence that radio-loud NLSy1 have $M_\bullet$ smaller than those of
blazars has recently been challenged by \cite{2013MNRAS.tmp..890C}.
Some nearby radio galaxies including Cen~A (NGC\,5128), 
Per~A (NGC\,1275, 3C\,84), and Vir~A (M87, 3C\,274) 
are detected by {\em Fermi}/LAT \citep{2010ApJ...720..912A}.
While part of their $\gamma$-ray luminosity is attributed 
to inverse-Compton scattering of CMB photons on
the extended (kpc-scale) radio lobes of the galaxies 
\citep{2007AIPC..921..325C,2010Sci...328..725A}, contribution
from the core region is also evident
\citep{2009ApJ...707...55A,2010ApJ...719.1433A}.
Unlike other radio galaxies studied by
\cite{2010ApJ...720..912A}, Per~A exhibits episodes of rapid GeV variability
\citep{2010ATel.2737....1D,2013ATel.4753....1C}.
The core $\gamma$-ray emission
in radio galaxies is probably produced by the same mechanisms as in blazars,
but with less extreme relativistic beaming.

Since early satellite observations established the association of some 
discrete $\gamma$-ray sources with AGNs, it became clear that 
blazars emit a considerable fraction of their total energy 
output above 100\,MeV
\citep{1978Natur.275..298S,1999ApJS..123...79H,2002BASI...30...73M}.
The current generation of space-based $\gamma$-ray telescopes that use
solid-state (silicon) detectors is represented by instruments onboard \agile{}
\citep{Tavani2009:AGILE,2008NIMPA.588...52T} and {\em Fermi}
\citep{2009ApJ...697.1071A}, which open a window into the world of GeV 
variability and spectral behavior of $\gamma$-ray-loud AGNs.
In contrast to previous expectations \citep{2004MNRAS.353..890V}, 
most of the brightest $\gamma$-ray blazars detected by {\em Fermi} and
\agile{} were already known from the EGRET era \citep{2011NIMPA.630....7T}.
On the other hand, many blazars previously unknown as $\gamma$-ray emitters
were observed to reach high fluxes ($>10^{-6}$\,photons\,cm$^{-2}$\,s$^{-1}$ at
energies $E>100$\,MeV) for only a short period of time during a flare.
In this work, we present a detailed investigation of one such object.

The radio source GB\,1310$+$487 [also known as GB6\,B1310$+$4844\footnote{The
correct B1950 source name, if its declination is expressed as three
digits, is 1310$+$487, while 1310$+$484 is an unrelated nearby radio source.}, 
and CGRaBS\,J1312$+$4828, listed in the {\em Fermi} $\gamma$-ray source 
catalogues as 1FGL\,J1312.4$+$4827 \citep{2010ApJS..188..405A} and 2FGL\,J1312.8$+$4828 
\citep{2012ApJS..199...31N}, radio VLBI position\footnote{see
\url{http://astrogeo.org/vlbi/solutions/rfc_2012b}} 
$\alpha_\mathrm{J2000} = 13\hr12\mm43\fsec353644 \pm 0.22$\,mas, 
$\delta_\mathrm{J2000} = +48\deg28\arcm30\farcs94047 \pm 0.16$\,mas 
\citep{2002ApJS..141...13B}] is a flat-spectrum radio source. 
It was unremarkable among other faint $\gamma$-ray detected blazars
(the $E>100$\,MeV flux during the first 11~months of the {\em Fermi} mission
was $\sim 3\times10^{-8}$\,photons\,cm$^{-2}$\,s$^{-1}$, as reported in the 1FGL
catalogue; \citealt{2010ApJS..188..405A}) until it appeared in the
daily {\em Fermi} sky with a flux of
$\sim 1.0\times10^{-6}$\,photons\,cm$^{-2}$\,s$^{-1}$ on 2009
November~18\footnote{UT dates are used through the text.}
\citep{2009ATel.2306....1S}.
\agile{} observations reported two days later confirmed the high-flux state 
of the source \citep{Bulgarelli2009:ATel1310}. Follow-up observations in the near-IR 
\citep{2009ATel.2311....1C} and optical \citep{2009ATel.2320....1I}
also found GB\,1310$+$487 in a high state compared to 
historical records. The daily average $\gamma$-ray
flux remained at $\sim 1.0\times10^{-6}$\,photons\,cm$^{-2}$\,s$^{-1}$ for
more than a week \citep{2009ATel.2316....1H}.

This paper presents multiwavelength observations of
GB\,1310$+$487 before, during, and after its active $\gamma$-ray state, and
suggests possible interpretations of the observed SED evolution. 
In Sect.~\ref{mwobse} we describe the observing techniques and data analysis.
Sect.~\ref{results} presents an overview of the observational results.
In Sect.~\ref{discussion} we discuss their implications, 
and we summarize our findings in Sect.~\ref{conclusions}.
Throughout this paper, we adopt the following convention: the spectral index
$\alpha$ is defined through the energy flux density as a function of
frequency $F_\nu \propto \nu^{+\alpha}$, the photon index $\Gamma_\mathrm{ph}$ 
is defined through the number of incoming photons as a function of energy 
$\mathrm{d}N(E)/\mathrm{d}E \propto E^{-\Gamma_\mathrm{ph}}$, and the two indices are related by
$\Gamma_\mathrm{ph}=1-\alpha$.
We use a $\Lambda$CDM cosmology, with the
following values for the cosmological parameters: $H_0 = 71$\,km\,s$^{-1}$\,Mpc$^{-1}$, 
$\Omega_\mathrm{m} = 0.27$, and $\Omega_{\Lambda} = 0.73$
\citep[see][]{2009ApJS..180..330K,1999astro.ph..5116H}, which corresponds to 
a luminosity distance of 
$D_L = 3800$\,$\mathrm{Mpc}$,
an angular-size distance of 
$D_A = 1400$\,$\mathrm{Mpc}$,
and a linear scale of 
6.9\,pc\,mas$^{-1}$ 
at the source redshift $z=0.638$
(see Sect.~\ref{sec:keckimgspec}).

\section{Multiwavelength observations}
\label{mwobse}

\subsection{Gamma-ray observations with Fermi/LAT}
\label{fermilat}

{\em Fermi} Gamma-ray Space Telescope (FGST; hereafter {\em Fermi}) is an orbiting observatory 
launched on 2008 June~11 by a Delta\,II rocket from
the Cape Canaveral Air Force Station in Florida, USA.
The main instrument aboard {\em Fermi} is the Large Area Telescope (LAT;
\citealt{2009ApJ...697.1071A,2009APh....32..193A,2012ApJS..203....4A}), 
a pair-conversion telescope designed to
cover the energy band from 20\,MeV to greater than 300\,GeV.
The {\em Fermi}/LAT is providing a unique combination of 
high sensitivity and a wide field of view of about $60^\circ$.
{\em Fermi} is operated in an all-sky survey mode most
of the time, which makes it ideal for monitoring AGN variability.

The dataset reported here was collected during
the first 33~months of {\em Fermi} science observations from 2008 August~4 to
2011 June~13 in the energy range 100\,MeV -- 100\,GeV.
The 33~month time interval is divided into subintervals according to the level of
 its $\gamma$-ray activity as observed by {\em Fermi}/LAT (see
Table~\ref{table:timeintervals}).

\begin{table*}
\caption{Changes in the $\gamma$-ray spectrum between time intervals considered in the analysis.}             
\label{table:timeintervals}      
{

\centering                          
      \begin{tabular}{rcccrr}
       \hline
        Period & UT interval & $100$\,MeV--$100$\,GeV flux & $\Gamma_\mathrm{ph}$ & TS & N \\
       \hline
        33 months & 2008-08-04 -- 2011-06-13 & $(1.03 \pm 0.04)\times10^{-7}$ & $2.18 \pm 0.02$ & 4415 & 3566 \\
        pre-flare & 2008-08-04 -- 2009-11-16 & $(0.34 \pm 0.05)\times10^{-7}$ & $2.41 \pm 0.09$ &  205 & 476  \\
        Flare\,1   & 2009-11-16 -- 2009-12-21 & $(6.94 \pm 0.32)\times10^{-7}$ & $1.97 \pm 0.03$ & 3333 & 947  \\
       Interflare& 2009-12-21 -- 2010-04-26 &  $(1.37 \pm 0.11)\times10^{-7}$ & $2.15 \pm 0.06$ &  917 & 592  \\
        Flare\,2   & 2010-04-26 -- 2010-07-26 & $(2.83 \pm 0.16)\times10^{-7}$ & $2.14 \pm 0.04$ & 1839 & 907  \\
       post-flare& 2010-07-26 -- 2011-06-13 & $(0.44 \pm 0.06)\times10^{-7}$ & $2.34 \pm 0.09$ &  236 & 422  \\
       \hline
      \end{tabular}

}

      {\bf Column designation:}
      Col.~1, $\gamma$-ray activity state;
      Col.~2, time interval used for spectral analysis;
      Col.~3, average flux in units of photons\,cm$^{-2}$\,s$^{-1}$;
      Col.~4, photon index: $\mathrm{d}N(E)/\mathrm{d}E \propto E^{-\Gamma_\mathrm{ph}}$;
      Col.~5, Test Statistic (TS) defined in Sect.~\ref{fermilat}; and
      Col.~6, number of photons attributed to the source (model dependent).
\end{table*}

{\em Fermi}/LAT data comprise a database containing arrival times,
directions, and energies of individual silicon-tracker events supplemented
by information about the spacecraft position and attitude needed to
calculate the effective exposure for a celestial region and time interval of interest.
The maximum-likelihood method is used to analyze these data by constructing
an optimal model of the sky region as a combination of point-like and diffuse sources
having a spectrum associated with each one of them
\citep{1996ApJ...461..396M,2010ApJS..188..405A}. 
The significance of source detection is quantified by the Test Statistic (TS) value, 
determined by taking twice the logarithm
of the likelihood ratio between the models including the target source 
($L_1$) and one including only the background sources ($L_0$): 
$\mathrm{TS} \equiv 2 (\ln L_1 - \ln L_0)$. The ratios $L_0$ and $L_1$ are maximized
with respect to the free parameters in the models.
The Monte-Carlo simulation performed by
\cite{1996ApJ...461..396M} for EGRET confirmed theoretical
predictions \citep{Wilks} that for a GeV telescope, in most cases, 
the TS distribution is close to $\chi^2$.

The unbinned likelihood analysis 
was performed with 
the \texttt{Fermi Science Tools}
package\footnote{For documentation of the \texttt{Science Tools}, see
\url{http://fermi.gsfc.nasa.gov/ssc/data/analysis/documentation/}} 
version v9r21p0. 
The \texttt{DIFFUSE} class events in the energy range 100\,MeV -- 100\,GeV were 
extracted from a region of interest defined as a circle of $15\deg$ 
radius centered at the radio position of GB\,1310$+$487. 
A cut on zenith angle $>100\deg$ was applied to reduce contamination 
from Earth-limb $\gamma$-rays, produced by cosmic rays interacting 
with the upper atmosphere \citep{2003A&A...398..391S,2009PhRvD..80l2004A}.
Observatory rocking angles of greater than $52\deg$ were also excluded.
A set of instrument response functions (IRFs) \texttt{P6\_V11\_DIFFUSE} was used
in the analysis. The sky model contained point sources from the 2FGL catalogue
\citep{2012ApJS..199...31N} within $20\deg$ from the target, as well as
Galactic \texttt{gll\_iem\_v02\_P6\_V11\_DIFFUSE.fit} and isotropic
\texttt{isotropic\_iem\_v02\_P6\_V11\_DIFFUSE.txt} diffuse
components\footnote{The models are available from the {\em Fermi} Science
Support Center
\url{http://fermi.gsfc.nasa.gov/ssc/data/access/lat/BackgroundModels.html}}. 
All point-source spectra were modeled with a power law; the photon 
index was fixed to the catalogue value for all sources except the target. 
The diffuse-background parameters were not fixed. 
The estimated systematic uncertainty of flux measurements with LAT 
using P6\_V11 IRFs is 10\% at 100\,MeV, 5\% at 500\,MeV, and 20\% at
10\,GeV and above\footnote{For newer P7\_V6 IRFs used in the adaptive lightcurve analysis
described below, the systematic uncertainties are lower at high energies:
10\% at 100\,MeV, 5\% at 560\,MeV, 10\% at 10\,GeV and above
\citep{2012ApJS..203....4A}; however, this difference is not critical for the
present analysis.}.

The lightcurve of the target source was constructed by applying the above 
analysis technique to a number of independent time bins.
The time bin width was chosen to be seven days.
Sources with less than one photon detected in the
individual bin or with TS $<25$ were excluded from the sky model for that bin.
The lightcurves were computed by integrating the power-law model in 
the energy range 100\,MeV -- 100\,GeV. 

For lightcurves with time bins of fixed widths,
the choice of bin width is a compromise between temporal
resolution and signal-to-noise ratio for the individual bins. 
For {\em Fermi}/LAT an alternative method has recently been developed
\citep{2012A&A...544A...6L},
in which the time bin widths are flexible and chosen
to produce bins with constant flux uncertainty.
Flux estimates are still produced with the standard
LAT analysis tools. 
In this case we used \texttt{Fermi Science Tools} v9r27p1 
and \texttt{P7CLEAN\_V6} event selection and
IRFs, for which the current version of the 
adaptive binning method has been optimized 
(we have checked that using the 
\texttt{P7SOURCE\_V6} class 
yields very similar fluxes).
At times of high source flux, the time bins are narrower
than during lower flux levels, therefore allowing us to study
more rapid variability during these periods.

The lower energy limit of the integral fluxes computed for the adaptively 
binned lightcurve is chosen to minimize the bin widths needed to reach 
the desired relative flux uncertainty for most bins. The derivation of 
this energy limit, called the optimum energy, is presented by
\cite{2012A&A...544A...6L}.
Because the source is variable and the optimum energy
value depends on the flux, we compute the optimum energy with
the average flux over the first two years of LAT
operation reported in the 2FGL catalogue \citep{2012ApJS..199...31N}.
The optimum energy is found to be $E_0=$ 283\,MeV for this source.
We produced two sets of adaptively binned lightcurves in the 283\,MeV --
200\,GeV energy range, one
with 25\% flux uncertainties and another with 15\% uncertainties.
For each of these uncertainty levels we created a second version
of the lightcurve by performing the adaptive binning in the
reverse-time direction.

\subsection{Gamma-ray observations with \agile{}/GRID}

The \agile{} \gray{} satellite \citep{Tavani2009:AGILE,2008NIMPA.588...52T} was launched 
on 2007 April~23 by a PSLV rocket from the Satish Dhawan Space Centre at
Sriharikota, India. 
\agile{} is a mission of the Italian Space Agency (ASI) devoted to high-energy
astrophysics, and is currently the only space mission capable of
observing cosmic sources simultaneously in the energy bands
18--60\,keV and 30\,MeV -- 30\,GeV thanks to its two scientific instruments: 
the hard X-ray Imager (Super-AGILE; \citealt{2007NIMPA.581..728F}) and the Gamma-Ray Imaging Detector
(GRID; \citealt{2009NIMPA.610..291R}). During the first two years of the
mission, \agile{} was mainly operated by performing 2--4~week-long
pointed observations, but following the reaction wheel malfunction in
October 2009 it was operated in a spinning observing mode, 
surveying a large fraction of the sky each day.

The \agile{}/GRID instrument detected enhanced \gray{} emission from \source{} from
2009 November~20 17:00 (JD\,2455156.2) to 2009 November~22 17:00 (JD\,2455158.2)
\citep[see ][for preliminary results]{Bulgarelli2009:ATel1310}.
Level~1 \agile/GRID data were reanalyzed using the \agile{} Standard
Analysis Pipeline \citep[see][for a description of
the \agile{} data reduction]{2009A&A...506.1563P,2010ApJ...712..405V}. We used \gray{} 
events from the \texttt{ASDCSTDe} archive, filtered by means of the
\texttt{FM3.119} pipeline.
Counts, exposure, and Galactic background \gray{} maps were created with
a bin size of $0.\!\!^{\circ}5 \times 0.\!\!^{\circ}5$\,, for $E \ge
100$\,MeV.
Since \agile{} was in its spinning observing mode, all maps were generated
including all events collected up to $50^{\circ}$ off-axis.
We rejected all \gray{} events whose reconstructed
directions form angles with the satellite-Earth vector smaller
than $90^{\circ}$,
reducing the \gray{} Earth limb contamination
by excluding regions within $\sim 20^{\circ}$ from the
Earth limb.
We used the latest version (\texttt{BUILD-20}) of the Calibration files
(\texttt{I0023}),
which will be publicly available at the ASI Science Data Centre
(ASDC) site\footnote{\url{http://agile.asdc.asi.it}},
and the \gray{} diffuse emission model \citep{Giuliani2004:diff_model}.
We subsequently ran the \agile{} Multi-Source Maximum Likelihood Analysis
(\texttt{ALIKE})  
task using a radius of analysis of 10$^{\circ}$ in order to obtain the
position and the flux
of the source.
A power-law spectrum with a photon index $\Gamma=2.1$ was assumed
in the analysis.

\subsection{X-ray observations with {\it Swift}/XRT}

The X-ray Telescope (XRT; \citealt{2005SSRv..120..165B}) onboard the {\it Swift} satellite
\citep{2004ApJ...611.1005G} provides simultaneous imaging and spectroscopic
capability over the 0.2--10\,keV energy range. 
The source GB\,1310$+$487 was observed by {\it Swift} at seven epochs during the two
$\gamma$-ray activity periods and in June 2011 during the low post-flare state.
A summary of the {\it Swift} observations is presented in
Table~\ref{table:swiftlog}. {\it Swift}/XRT was operated in photon-counting 
(pc) mode during all observations. The low count~rate of the source 
allows us to neglect the pile-up effect which is
of concern for the XRT in pc mode if the count~rate\footnote{\url{http://www.swift.ac.uk/pileup.shtml}} is $\ge
0.6$\,count\,s$^{-1}$.

The \texttt{xrtpipeline} task from the \texttt{HEASoft v6.14} package was
used for the data processing with the standard filtering criteria.
To increase the number of counts for spectral analysis, the resulting event 
files were combined with \texttt{xselect} to produce average X-ray spectra
for the periods of Flare\,1 and Flare\,2 defined in
Table~\ref{table:timeintervals}. 
The spectrum for the Flare\,1 period was binned 
to contain at least 25 counts per bin to utilize the $\chi^2$
minimization technique.
The combined spectra were analyzed with
\texttt{XSPEC v12.8.1}. The simple absorbed power-law model with the
H~I column density fixed to the Galactic value $N_{\rm H~I} = 0.917 \times 10^{20}$\,cm$^{-2}$
(obtained from radio 21\,cm measurements by
\citealt{2005A&A...440..775K}) provided a statistically acceptable fit
(reduced $\chi^2 = 1.2$ for 12 degrees of freedom) to the 0.3--10\,keV spectrum. 
Leaving $N_{\rm H~I}$ free to vary results in the
values $N_{\rm H~I} =0.4{\rm-}0.5 \times 10^{22}$\,cm$^{-2}$ and
$\Gamma_\mathrm{ph~X-ray} \sim 1.3$. However, this model does not improve the fits. 
The low photon counts prevent a more detailed study.

Individual observations obtained 
during the periods of Flare\,1 and Flare\,2 
were also analyzed using the same fixed--$N_{\rm H~I}$ model, but no
evidence of spectral variability within the periods was found; however, the
low photon counts could easily hide mild spectral changes.
The {\it Swift}/XRT observation obtained during the Flare\,2 and post-flare
intervals (Table~\ref{table:timeintervals}) resulted in a low number of detected
photons. The \cite{1979ApJ...228..939C} statistic is applied to fit this
dataset with the absorbed power-law model. The Cash statistic is based on a likelihood ratio test and is
widely used for parameter estimation in photon-counting experiments.
The net count rate in the 0.3--10 keV energy range changed by a factor of 1.6 between the two 
observations conducted during Flare\,1 and by a factor of 3.5 over the whole
33-month period.
The X-ray spectral analysis results are presented in Table~\ref{table:swiftlog}.

\begin{table*}
\caption{{\it Swift} observations of GB\,1310$+$487.}             
\label{table:swiftlog}      
{
\centering                          
\begin{tabular}{cccccccc}        
\hline\hline                 
ObsID &  Date (UT)      & JD     & Exposure & UVOT                            & XRT        & 0.3--10\,keV unabs. flux                & $\Gamma_\mathrm{ph~X-ray}$  \\    
      &        &2455... &  (ks)    &  (mag)                        & (cts/s)    & ($10^{-13}$\,erg\,cm$^{-2}$\,s$^{-1}$)   &   \\
\hline                         
\multicolumn{8}{l}{Flare\,1 period}\\                                       
 001  & 2009-11-27 & 162.77 & 8.7      & $M2=21.1(2)$                    & $0.022(2)$ & \multirow{2}{*}{$16.7 \pm 3.3$} & \multirow{2}{*}{$0.88 \pm 0.18$}   \\      
 002  & 2009-11-30 & 165.97 & 8.1      & $U=21.2(4)$, $B>21.3$, $V>20.9$ & $0.013(1)$ &                                 &    \\ 
\hline
\multicolumn{8}{l}{Flare\,2 period}\\
 003  & 2010-06-25 & 372.66 & 4.5      & $U>21.3$, $B>21.5$, $V>20.5$    & $0.010(1)$ & \multirow{4}{*}{$6.6 \pm 1.5$} & \multirow{4}{*}{$1.15 \pm 0.23$}    \\
 004  & 2010-07-03 & 380.65 & 4.5      & $U=20.9(3)$, $B>21.5$, $V>20.3$ & $0.009(1)$ &                                &    \\ 
 005  & 2010-07-07 & 384.63 & 4.9      & $U=21.4(3)$                     & $0.009(1)$ &                                &    \\
 006  & 2010-07-11 & 388.68 & 4.6      & $U=21.3(3)$                     & $0.010(1)$ &                                &    \\
\hline
\multicolumn{8}{l}{post-flare period}\\
 007  & 2011-06-03 & 717.12 & 9.5      & $U=21.1(2)$                     & $0.006(1)$ & $5.5 \pm 1.8$                  & $0.93 \pm 0.34$   \\ 
\hline                                   
\end{tabular}

}

{\bf Column designation:}
Col.~1, observation number in the {\it Swift} archive omitting the leading 00031547;
Cols.~2,~3, date of observation given by the Gregorian and Julian Date;
Col.~4, exposure time in kiloseconds;
Col.~5, {\it Swift}/UVOT photometry (here and later in the text the
error in parentheses corresponds to the last decimal place of the value
before the parentheses);
Col.~6, {\it Swift}/XRT net count~rate in counts/s and its uncertainty;
Col.~7, 0.3--10\,keV unabsorbed flux derived from fitting {\it Swift}/XRT
data with the power-law model (datasets 1--2 and 3--6 are combined to
increase the photon statistics); and
Col.~8, X-ray spectral index ($\Gamma_\mathrm{ph~X-ray}$) as defined in
Sect.~\ref{intro}.

\end{table*}

\subsection{Ultraviolet--optical observations}
\label{uvoptobs}


The {\em Swift} Ultraviolet-Optical Telescope (UVOT;
\citealt{2005SSRv..120...95R}) 
has a diameter of 0.3\,m 
and is equipped 
with a microchannel-plate intensified CCD detector operated in photon-counting mode.
{\it Swift}/UVOT observed GB\,1310$+$487 simultaneously with 
{\it Swift}/XRT. 
Various filters were used at different epochs ranging from the $U$ to $M2$ bands
(as detailed in Table~\ref{table:swiftlog}), with the best coverage achieved
in the $U$ band. 
Since the object is very faint, multiple subexposures taken during each
observation were stacked together with the tool \texttt{uvotimsum} from the
\texttt{HEASoft} package. A custom-made script based on \texttt{uvotsource} 
was employed for 
aperture photometry (using the standard $5\arcs$ aperture diameter) and count
rate~to~magnitude
conversion taking into account the coincidence loss (pile-up) correction
\citep{2008MNRAS.383..627P,2010MNRAS.406.1687B}.
The Galactic reddening in the direction of this source is
$E(B-V)=0.013$\,mag 
\citep{1998ApJ...500..525S}. Using the extinction law of
\cite{1989ApJ...345..245C} and coefficients presented by 
\cite{2009ApJ...690..163R}, the following extinction values 
were obtained for the individual bands: 
$A_V = 0.041$,
$A_B = 0.053$,
$A_U = 0.065$,
and $A_{M2}= 0.122$ mag.
Magnitude-to-flux-density conversion was performed using the calibration
of \cite{2008MNRAS.383..627P}.

A star-like object is visible in Nordic Optical Telescope (NOT) images about $3\arcs$
southwest of the AGN. This object would be blended with the AGN in
UVOT images which lack adequate angular resolution. Contribution of this object to
the total flux measured by UVOT is the likely reason for the discrepancy between
UVOT and NOT $U$-band measurements during the low state of GB\,1310$+$487.
The nearby galaxy (Sect.~\ref{results}) also contributes to the measured UVOT flux.

\begin{figure}
\centering
\includegraphics[width=0.48\textwidth]{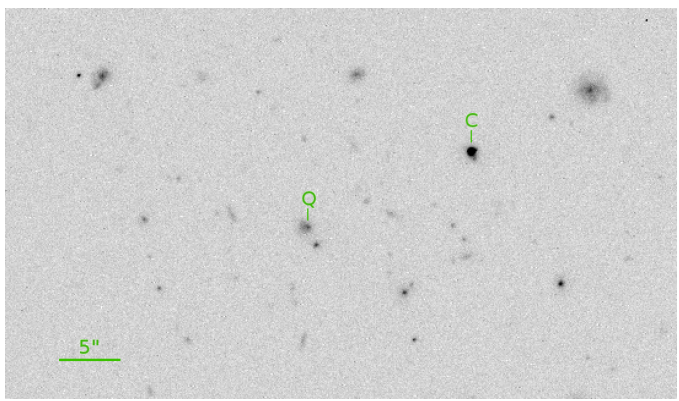}
\caption {Nordic Optical Telescope $R$-band image of the GB\,1310$+$487 
region obtained on 2011 May~29. The exposure time was 300\,s. 
North is up and east is to the left. The AGN and comparison star
SDSS\,J131240.83$+$482842.9 used in the NOT data analysis are marked with
the letters ``Q'' and ``C,'' respectively.}
\label{figure:notr}
\end{figure}

The Nordic Optical Telescope, a 2.5\,m instrument 
located on La~Palma, Canary Islands, conducted photometric observations of GB\,1310$+$487 
with its ALFOSC camera on 2010 July~7 and 2011 May~29 during the second flare and the 
post-flare low state, respectively. The
VaST\footnote{\url{http://scan.sai.msu.ru/vast}} software
\citep{2005ysc..conf...79S} was applied for
the basic reduction (bias removal, flat-fielding) and aperture photometry of
the NOT images. A fixed aperture $1\farcs5$ in diameter was used for the
measurements. 
The source 3UC\,277-116569, which served as the comparison star for the Kanata observations
(see below), 
was saturated on NOT $i$~band images and so could not be used.
Instead, SDSS\,J131240.83$+$482842.9 ($\alpha_\mathrm{J2000} = 13\hr12\mm40\fsec84$, 
$\delta_\mathrm{J2000} = +48\deg28\arcm42\farcs9$,
\citealt{2009ApJS..182..543A}; see Fig.~\ref{figure:notr}) was used as the comparison star. Its
Johnson-Cousins magnitudes were computed from the SDSS photometry using
conversion formulas of \cite{2006A&A...460..339J}:
$U= 19.133 \pm 0.081$,
$B= 19.303 \pm 0.013$,
$V= 18.822 \pm 0.012$,
$R= 18.589 \pm 0.011$, and
$I= 18.177 \pm 0.018$ mag.

Kanata, a 1.5\,m telescope at the Higashi-Hiroshima Observatory, observed 
GB\,1310$+$487 in the $R$ and $I$ bands with the HOWPol instrument \citep{2008SPIE.7014E.151K} 
in the nonpolarimetric mode for eight nights during the first 
and second $\gamma$-ray flares. Relative point-spread function (PSF) photometry was conducted using 
3UC\,277-116569 ($\alpha_\mathrm{J2000} = 13\hr12\mm54\fsec09$,
$\delta_\mathrm{J2000} = +48\deg27\arcm58\farcs2$, J2000; $R = 16.109$, $I = 15.657$ mag;
\citealt{2010AJ....139.2184Z}) as the comparison star.
The adopted Galactic extinction values were 
$A_R=0.035$ and
$A_I=0.025$ mag
\citep{1998ApJ...500..525S}.
The calibration by \cite{1998A&A...333..231B} was employed for the magnitude-to-flux conversion.


The source GB\,1310$+$487 was assigned a redshift of 0.501 based on
a 2007 March~21 1200\,s Hobby-Eberly Telescope Low Resolution Spectrograph
(HET/LRS) observation\footnote{\cite{1998ApJ...494...47F} 
previously reported $z=0.313$, but it was indicated as a ``marginal measurement.''},
which showed strong [O~II]~$\lambda$3727 at 5592\,$\AA$  and weak evidence
of host absorption features \citep{2008ApJS..175...97H,2012ApJ...748...49S}. 
This spectrum had insufficient signal-to-noise ratio (S/N) to exclude weak broad lines,
or to cleanly measure the optical continuum, leaving the nature
of the source uncertain.

\begin{figure}
\centering
\includegraphics[width=0.48\textwidth]{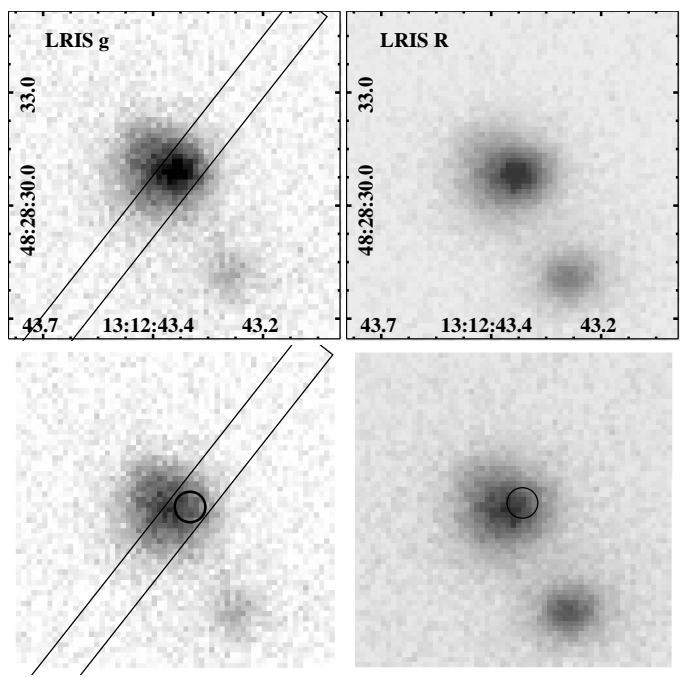}
\caption{ Keck\,I/LRIS images of GB\,1310$+$487
obtained on 2013 May~10. 
Top panel: the $g$- and $R$-band images ($8\arcs\times8\arcs$
field of view; north is up and east is to the left).
Bottom panel: $g$- and $R$-band images after subtraction of a scaled point source from the offset
core. The boxes show the location of the DEIMOS slit in
the 2013 April~07 observations. The circles indicate
the location of the radio-loud AGN. These residual images reveal a relatively
undisturbed foreground galaxy.}
\label{figure:keckimg}
\end{figure}

Thus, we reobserved the source
with the Keck 10\,m telescopes. Long-slit spectra were
obtained with the Keck\,II DEep Imaging Multi-Object Spectrograph (DEIMOS) \citep{2003SPIE.4841.1657F} on 2013 April~07
with $\sim 1''$ seeing. Two 600\,s
integrations were obtained using the 600~lines~per~mm (7500\,$\AA$
~blaze\footnote{The ``blaze wavelength'' is the wavelength for which
the grating is the most efficient.}) grating,
providing coverage in the 4450--9635\,\AA\ range with a $\sim 100$\,\AA\ gap
between the two CCDs. With the $1\farcs0$ slit, 
the spectra have an effective resolution of $\sim 3.0$\,\AA. Conditions
were good, but not completely photometric; the flux scale might be 
uncertain by roughly a factor of 2.
Moreover, we obtained $2 \times 180$\,s 
$g$- and $R$-band images of the object with the two cameras on the Keck\,I 
Low Resolution Imaging Spectrometer (LRIS)
\citep{1995PASP..107..375O}
on May~10 under $\sim 1''$ seeing, as shown in Figure~\ref{figure:keckimg}.
The DEIMOS slit on April~07 was placed on the bright core of the source,
at the parallactic angle 
\citep{1982PASP...94..715F} of $\mathrm{PA}=143^\circ$ (measured from N to E), and the extended wings of the host were also included.
The companion was $\sim 3\arcs$ off the slit.

A second Keck\,II/DEIMOS spectrum was obtained on 
June~10 with a different slit position and $\sim 0.7''$ seeing.
It has a higher signal-to-noise ratio than the first DEIMOS spectrum; however, it
was affected by a cosmic-ray hit that prevented 
accurate measurement of H$\gamma$ in the $z=0.638$ system, and
conditions were not photometric when the standard star was being
observed.
The two spectra are normalized to epoch~1 (April~07) using
the [O~II]~$\lambda$3727 line at $z=0.500$.
The continuum cannot be used to cross-calibrate the two spectra because
of the significantly variable AGN flux contribution; the second-epoch
continuum level appears to have dropped relative to the emission lines
by $\sim 1/3$. The two spectra were
averaged for further analysis.

\subsection{Infrared photometry}
\label{sec:irphotometry}

Observations in the near-IR were carried out with the 2.1\,m 
telescope of the Guillermo Haro Observatory, INAOE, Mexico.
The telescope is equipped with the CANICA camera 
together with $J$, $H$, and $K_s$ filters. 
We carried out differential photometry between the object of interest
and other objects in the $5\arcm\times5\arcm$ field.  
The observations showed an increase of about 
one magnitude during the Flare\,2 period with respect to the Flare\,1 and post-flare
periods (results are summarized in Table\,\ref{table:photometry}).
Magnitudes are referred
to the 2MASS\footnote{\url{http://www.ipac.caltech.edu/2mass/}} survey published photometry.
The source GB\,1310$+$487 itself was not detected in the 2MASS survey. The
survey detection limit is $J=15.8$, $H=15.1$, and $K_s=14.3$ mag
\citep{2006AJ....131.1163S}. These values may be considered upper limits on
the object brightness at the 2MASS observation epoch of JD\,2451248.8408
(1999 March 11).

The source GB\,1310$+$487 is listed in the Wide-field Infrared Survey Explorer (WISE;
\citealt{2010AJ....140.1868W}) catalogue \citep{2012wise.rept....1C} with the
following magnitudes in the four WISE bands: 3.4\,$\mu$m W1 $=12.302\pm0.024$,
4.6\,$\mu$m W2 $=11.254\pm0.021$, 12\,$\mu$m W3 $=8.596\pm0.021$, and 22\,$\mu$m
W4 $=6.368\pm0.044$. The IR colors (W1--W2 $=1.048\pm0.032$,
W2--W3 $=2.658\pm0.030$ mag) are at the blue edge of the area in the color--color 
diagram occupied by blazars and Seyfert galaxies (see Fig.~12 in
\citealt{2010AJ....140.1868W}, Fig.~1 in \citealt{2012ApJ...748...68D}, 
and Fig.~1 in \citealt{2011ApJ...740L..48M}), indicating that the AGN and not
the host galaxy's stars or warm dust is responsible for most of the IR flux in these bands.
WISE observations of this area were conducted on 2010 June~3--8 during
Flare\,2 (Table~\ref{table:timeintervals}).

\begin{table}
\caption{Ground-based photometry of GB\,1310$+$487.}             
\label{table:photometry}      
{
\centering                          
\begin{tabular}{ccclc}        
\hline\hline                 
Date       & JD (UTC)& Filter    & mag     & Instrument  \\    
           &2455... &           &          &   \\
\hline                         
\multicolumn{5}{l}{Flare\,1 period}\\                                       
2009-11-28 & 164.33668 & $R$ &    20.58(2) & Kanata \\
2009-11-29 & 165.33620 & $R$ &    20.86(9) & Kanata \\
2009-12-05 & 171.31613 & $R$ &    20.61(8) & Kanata \\

2009-11-28 & 164.34442 & $I$ &    19.41(2) & Kanata \\
2009-11-29 & 165.35134 & $I$ &    19.66(6) & Kanata \\
2009-12-05 & 171.31613 & $I$ & $>$18.91    & Kanata \\
2009-12-13 & 179.33910 & $I$ &    19.52(1) & Kanata \\

2009-11-22 & 158.04390 & $H$ & 15.87(6)    & OAGH \\

\multicolumn{5}{l}{interflare period}\\
2010-03-17 & 272.95922 & $H$ & 15.97(5)    & OAGH \\

\multicolumn{5}{l}{Flare\,2 period}\\
2010-07-07 & 385.44392 & $U$ & 21.91(6)    & NOT \\
2010-07-07 & 385.47325 & $B$ & 22.6(1)     & NOT \\
2010-07-07 & 385.43551 & $V$ & 21.62(5)    & NOT \\

2010-06-03 & 351.03843 & $R$ & $>$20.78    & Kanata \\
2010-06-04 & 352.05203 & $R$ & $>$20.52    & Kanata \\
2010-06-05 & 353.02920 & $R$ & $>$20.64    & Kanata \\
2010-07-07 & 385.47773 & $R$ & 20.85(2)    & NOT \\
2010-07-17 & 395.07192 & $R$ & $>$19.90    & Kanata \\
2010-07-19 & 396.99981 & $R$ & $>$20.48    & Kanata \\

2010-06-03 & 351.04981 & $I$ &    19.98(7) & Kanata \\
2010-06-05 & 353.04057 & $I$ &    19.80(2) & Kanata \\
2010-07-07 & 385.46087 & $I$ & 19.79(2)    & NOT \\
2010-07-17 & 395.08708 & $I$ & $>$19.27    & Kanata \\

2010-06-15 & 362.70970 & $J$ & 16.4(1)     & OAGH \\
2010-06-16 & 363.75472 & $J$ & 16.6(1)     & OAGH \\
2010-06-18 & 365.69752 & $J$ & 16.3(1)     & OAGH \\
2010-06-19 & 366.71342 & $J$ & 15.7(1)     & OAGH \\

2010-05-17 & 333.79362 & $H$ & 14.59(1)    & OAGH \\
2010-05-20 & 336.83470 & $H$ & 15.11(5)    & OAGH \\
2010-06-15 & 362.70186 & $H$ & 14.61(7)    & OAGH \\
2010-06-16 & 363.74752 & $H$ & 14.68(3)    & OAGH \\
2010-06-19 & 366.69745 & $H$ & 14.84(5)    & OAGH \\

2010-06-15 & 362.72145 & $K_s$ & 13.8(1)   & OAGH \\
2010-06-16 & 363.76233 & $K_s$ & 13.78(8)  & OAGH \\
2010-06-19 & 366.72800 & $K_s$ & 13.7(1)   & OAGH \\

\multicolumn{5}{l}{post-flare period}\\
2011-05-29 & 711.46544 & $V$ & 21.56(4)    & NOT \\
2011-05-29 & 711.46135 & $R$ & 20.79(2)    & NOT \\
2011-05-29 & 711.46922 & $I$ & 19.87(3)    & NOT \\
		
2011-07-31 & 773.69450 & $H$ & 15.83(7)    & OAGH \\
\hline                                   
\end{tabular}

}

{\bf Column designation:}
Cols.~1,~2, the Gregorian and Julian Date of observation, respectively;
Col.~3, filter;
Col.~4, magnitude and its uncertainty; and
Col.~5, telescope name.

\end{table}

\subsection{Radio observations}

As part of an ongoing blazar monitoring program, the Owens Valley
Radio Observatory (OVRO) 40\,m radio telescope has observed GB\,1310$+$487
at 15\,GHz regularly since the end of
2007~\citep{richards_et_al_2011}. This monitoring program studies
over 1500 known and likely $\gamma$-ray-loud blazars, including all
CGRaBS~\citep{2008ApJS..175...97H} sources north of declination $-20^{\circ}$.  
The objects in this program are observed in total intensity twice per week.
The minimum measurement uncertainty is 4\,mJy while the typical uncertainty
is 3\% of the measured flux.
Observations are performed
with a dual-beam (each 2.5$\arcm$ full width at half-maximum intensity, FWHM) Dicke-switched system using
cold sky in the off-source beam as the reference. Additionally, the
source is switched between beams to reduce atmospheric variations. The
absolute flux-density scale is calibrated using observations of
3C\,286, adopting the flux density (3.44\,Jy) from
\citet{1977A&A....61...99B}. This results in a $\sim 5$\% absolute
flux-density-scale uncertainty, which is not reflected in the plotted errors.

Multifrequency radio observations of GB\,1310$+$487 were performed with
the 100\,m~Effelsberg telescope operated by the MPIfR\footnote{Max-Planck-Institut f\"ur Radioastronomie}. 
Observations were conducted on 2009 December~1 following the reported $\gamma$-ray flare,
on 2010 June~28 during the second $\gamma$-ray active phase, and on 2011 June~5. Secondary focus 
heterodyne receivers operating at 2.64, 4.85, 8.35, 10.45, and 14.60\,GHz were
used. The observations were conducted with cross-scans (i.e., the
telescope's response was measured while slewing over the source position in azimuth and
elevation). The measurements were corrected for
(a)~pointing offsets, (b)~atmospheric opacity, and (c)~elevation-dependent gain 
(see \citealt{2008A&A...490.1019F,2009A&A...501..801A}).
The multifrequency observations were completed within 1\,hr for each observing
session. 
The absolute flux-density calibration was done by observing standard
calibrators such as, 3C\,48, 3C\,161, 3C\,286, 3C\,295, and NGC\,7027 
\citep{1977A&A....61...99B,1994A&A...284..331O}.

The IRAM~30\,m Pico Veleta telescope observations at
86.24 and 142.33\,GHz took place on 2009 December~7. The observations 
and data-reduction strategy were similar to those with Effelsberg; 
a detailed description is given by \cite{Nestorassub}. Both the
Effelsberg~100\,m and IRAM~30\,m telescope observations were conducted in the
framework of the F-GAMMA program 
\citep{2007AIPC..921..249F,2008MmSAI..79.1042A,2010arXiv1006.5610A,2012JPhCS.372a2007A,Fuhrmannsub}.

For comparison with the latest Effelsberg, IRAM, and OVRO results, we use data from
the RATAN-600 576\,m ring radio telescope of the Special Astrophysical Observatory 
(Russian Academy of Sciences); RATAN-600 observations of GB\,1310$+$487 
were performed in transit mode at the southern sector with the flat reflector
quasi-simultaneously at 3.9, 7.7, 11.1, and 21.7\,GHz in June 2003 within the
framework of the spectral survey conducted by 
\cite{1999IAUS..194..177K, 2002PASA...19...83K}. The flux-density scale 
is set using calibrators listed by \cite{1977A&A....61...99B,1994A&A...284..331O}.
Details on the RATAN-600 observations and data processing are discussed 
by \cite{1999A&AS..139..545K}.

The National Radio Astronomy Observatory's Very Long Baseline Array (VLBA,
\citealt{1994IAUS..158..117N,1995ASPC...82...59N}) is a system of ten 
25\,m radio telescopes dedicated to very long baseline interferometry (VLBI) 
observations for astrophysics, astrometry, and geodesy. 
After publication of the report on the November 2009 $\gamma$-ray flare
\citep{2009ATel.2306....1S}, GB\,1310$+$487 was added to the 
MOJAVE\footnote{Monitoring Of Jets in Active galactic nuclei with VLBA Experiments, \url{http://www.physics.purdue.edu/astro/MOJAVE/}}
program \citep{2009AJ....137.3718L}. 
Three epochs of VLBA
observations at 15\,GHz were obtained between 2009 and 2010.

\section{Results}
\label{results}

\subsection{$\gamma$-ray analysis}
\label{sec:grayresults}

The $\gamma$-ray counterpart of GB\,1310$+$487 was localized, integrating
33~months of {\em Fermi}/LAT monitoring data, to 
$\alpha_\mathrm{J2000} = 198\fdeg187$,
$\delta_\mathrm{J2000} = 48\fdeg472$, with a 68\% uncertainty of $0\fdeg014 = 50\arcs$. 
This is a factor of five larger than the spacecraft alignment accuracy of
$10\arcs=0\fdeg003$ \citep{2012ApJS..199...31N}.
The $\gamma$-ray position is only $0\fdeg005 = 18\arcs$ away from the radio position of
GB\,1310$+$487. 
Within the Fermi/LAT error circle no other radio sources are seen with the
VLA FIRST 1.4\,GHz survey \citep{1997ApJ...475..479W}, which provides
the best combination of sensitivity
and angular resolution for that region of the radio sky to date.
Therefore, the positional association of the $\gamma$-ray source with the
radio source GB\,1310$+$487 is firmly established.
The X-ray brightening observed during the first 
and, to a lesser extent, the second $\gamma$-ray flares 
together with the near-IR brightening during the second $\gamma$-ray flare
support the identification of the $\gamma$-ray source with 
the lower-frequency counterpart. 
See
Tables~\ref{table:timeintervals},~\ref{table:swiftlog},~\ref{table:photometry},
and the discussion below for details.

\begin{figure}
 \centering
 \includegraphics[width=0.5\textwidth,angle=0]{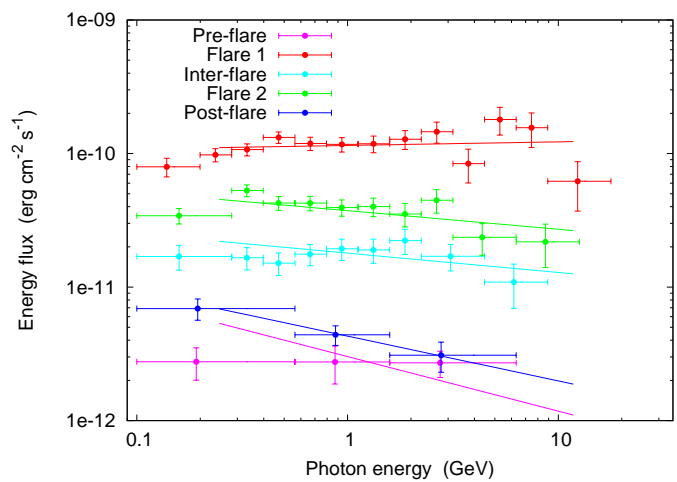}
 \caption{{\em Fermi}/LAT $\gamma$-ray SEDs compared to the
 power-law models (shown as lines in this logarithmic plot) 
 derived from unbinned likelihood analysis. We stress that the power-law
 models are {\em not} fits to the binned energy flux values; rather, these 
are two independent ways to represent {\it Fermi}/LAT photon data.
The time intervals defined in Table~\ref{table:timeintervals} are color
labeled.}
 \label{figure:gs}
\end{figure}

The $\gamma$-ray spectra of GB\,1310$+$487 at various activity states listed in
Table~\ref{table:timeintervals} are presented in Figure~\ref{figure:gs}.
The plotted spectral bins satisfy the following requirements: 
$\mathrm{TS}>50$  and/or model-predicted number of source photons $N>8$. 
The flux value at each bin was computed by fitting the model with
source position and power-law photon index fixed to the values estimated 
over the entire period in the 0.1--100\,GeV energy range. The Galactic component 
parameters were fixed, as were all other nontarget source components of the model.

To test if the power law (PL) is an adequate approximation of the observed
$\gamma$-ray spectrum, the PL fit (represented by a straight line if
plotted on a logarithmic scale) was compared to the fit with a 
log-parabola (LP) function 
defined as $\mathrm{d}N/\mathrm{d}E=N_0(E/E_0)^{-(\alpha+\beta \ln(E/E_0))}$, 
where $N$ is the number of photons with energy $E$,
$N_0$ is the normalization coefficient,
$E_0$ is a reference energy, $\alpha$ is the spectral slope at energy $E_0$,
and $\beta$ is the curvature parameter around the peak. 
For the combined 33~month {\em Fermi}/LAT
dataset, the fit with the assumption of a PL spectrum for the target source
provides its detection with the Test Statistic $\mathrm{TS}_\mathrm{PL}=4415$, while the LP spectrum leads
to $\mathrm{TS}_\mathrm{LP}=4423$. These values may be compared by defining,
following \cite{2012ApJS..199...31N} and in analogy with the source detection TS described in
Sect.~\ref{fermilat},
the curvature Test Statistic 
$\mathrm{TS}_\mathrm{curve} \equiv 2 (\ln L_\mathrm{LP} - \ln L_\mathrm{PL})
= \mathrm{TS}_\mathrm{LP} - \mathrm{TS}_\mathrm{PL}$. The obtained value of
$\mathrm{TS}_\mathrm{curve} = 8$ corresponds to a $2.8 \sigma$ difference,
which is lower than the $\mathrm{TS}_\mathrm{curve} > 16$ ($4 \sigma$)
threshold applied by \cite{2012ApJS..199...31N}. We conclude that while there is a hint of
spectral curvature, it cannot be considered significant. 

The broken power-law (BPL) model was also tested, but it did not provide a
statistically significant improvement over the PL or the LP
models ($\mathrm{TS}_\mathrm{BPL} = 4423$, for the best-fit break energy
$E_b=3$\,GeV, the photon indexes $\Gamma_{\mathrm{ph}\,1}=2.30\pm0.04$,
$\Gamma_{\mathrm{ph}\,1}=0.05\pm0.02$ above and below the
break, respectively).
Therefore, we adopt the simpler PL model for the following analysis.

\begin{figure}[!htb]
 \centering
 \includegraphics[width=0.5\textwidth]{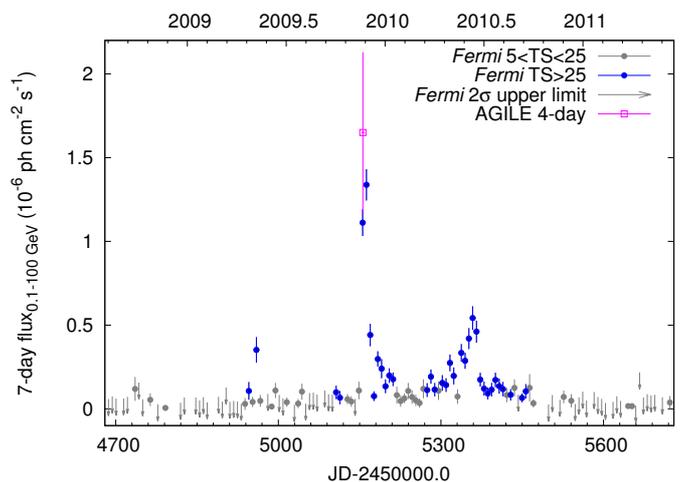}
 \caption{Weekly binned {\it Fermi}/LAT lightcurve. Blue filled circles are values with
$\mathrm{TS}>25$, gray filled circles are values with $5<\mathrm{TS}<25$, and gray arrows
indicate $2\sigma$ upper limits for time bins with no significant detections
($\mathrm{TS}<5$). A four-day integrated {\it AGILE} data point is added as an open box for comparison.}
 \label{figure:lightcurve}
\end{figure}

The 33-month $\gamma$-ray lightcurve of the source obtained with the seven-day
binning is presented in Figure~\ref{figure:lightcurve}. Two major flaring
periods are clearly visible. The first, brighter flare peaked around 
2009 November~27 (JD\,2455163) with the weekly averaged flux of 
$(1.4 \pm 0.1)\times10^{-6}$\,photons\,cm$^{-2}$\,s$^{-1}$. The peak flux 
averaged over the two-day interval centered on that date is 
$(1.9 \pm 0.2)\times10^{-6}$\,photons\,cm$^{-2}$\,s$^{-1}$.
The source continued to be observed at a daily flux of 
$\sim 0.5 \times 10^{-6}$\,photons\,cm$^{-2}$\,s$^{-1}$ for another two weeks.
The second flare peaked around 2010 June~17 (JD\,2455365) at the seven-day integrated flux of 
$(0.54 \pm 0.07)\times10^{-6}$\,photons\,cm$^{-2}$\,s$^{-1}$. The daily flux
of $\sim 0.5 \times10^{-6}$\,photons\,cm$^{-2}$\,s$^{-1}$ was observed for
about three weeks around this date.
The two flares demonstrate remarkably contrasting flux evolution: the first
is characterized by a fast rise and slower decay, while the second flare 
shows a gradual flux rise followed by a sharp decay. 
Following \cite{1974ApJ...193...43B}, \cite{1999ApJS..120...95V}, and \cite{2008ARep...52..278G}, we
define the flux-variability timescale as 
$t_\mathrm{var} \equiv \Delta t / \Delta \ln{S} $,
where $\Delta \ln{S}$ is the difference
in logarithm of the photon flux at two epochs separated by the time interval
$\Delta t$.
The observed flux-variability timescale during the onset of Flare\,1,
as estimated from the seven-day binned lightcurve (Fig.~\ref{figure:lightcurve}), is
$t_\mathrm{var} \approx 3$~days. 
The timescale of flux decay after Flare\,2 is $t_\mathrm{var} \approx
5$~days. 

\begin{figure}[!htb]
 \centering
 \includegraphics[width=0.5\textwidth,angle=0]{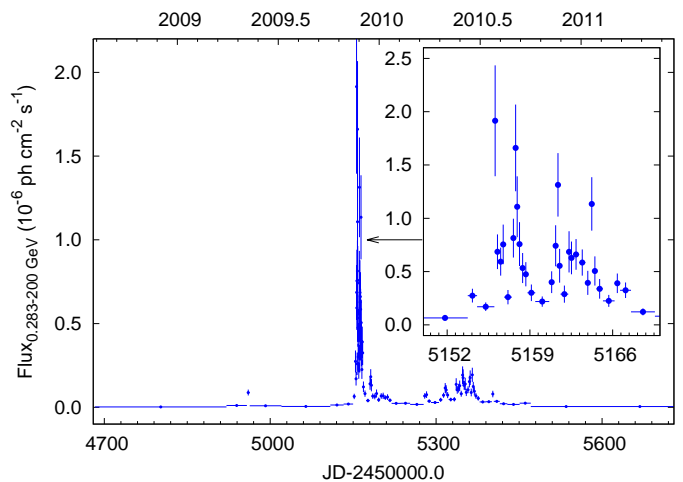}
 \caption{{\em Fermi}/LAT lightcurve constructed with the adaptive binning method
 \citep{2012A&A...544A...6L}. The magnified plot of Flare\,1 is shown in the insert.
The energy range for this lightcurve is chosen to minimize the
uncertainties in time and flux, while the lightcurve was 
Fig.~\ref{figure:lightcurve} is given in the commonly used $E > 100$\,MeV energy range.}
 \label{figure:lightcurveadaptive}
\end{figure}

The {\em Fermi}/LAT lightcurve constructed with the alternative
analysis method, the adaptive binning (with 25\% flux uncertainty
at each bin), is presented in Figure~\ref{figure:lightcurveadaptive}.
It confirms all the features visible in the constant bin-width lightcurve, 
but also allows us to investigate fast variability during high-flux states in greater detail.
The first flare episode, Flare\,1, 
consists of four prominent subflares, each with a time width of a day or
less. The subflares show no obvious asymmetry and the
variability timescale $t_\mathrm{var}$ for the three point rise of the
second and third subflares (at JD\,2455157.5 and 2455161.5)
was estimated as $0.36 \pm 0.20$~days and $0.45 \pm 0.23$~days, respectively.
A second adaptively binned lightcurve was produced in the reverse-time 
direction, which gives a similar, but not identical, time binning.
The result of the timescale estimates for this second version of the
lightcurve was found to be consistent with the first analysis.
A similar analysis for the lightcurves with 15\% uncertainties
give timescale estimates of about 1~day for the most rapid
variability. We conclude that the adaptively binned lightcurves
show evidence of a variability timescale of half a day with a
conservative upper limit of 1~day. For the second and
fainter flare epoch the timescales seen in the adaptive binning
are consistent with the estimate from the fixed-binned lightcurve 
described above.

\begin{figure}
 \centering
 \includegraphics[width=0.5\textwidth]{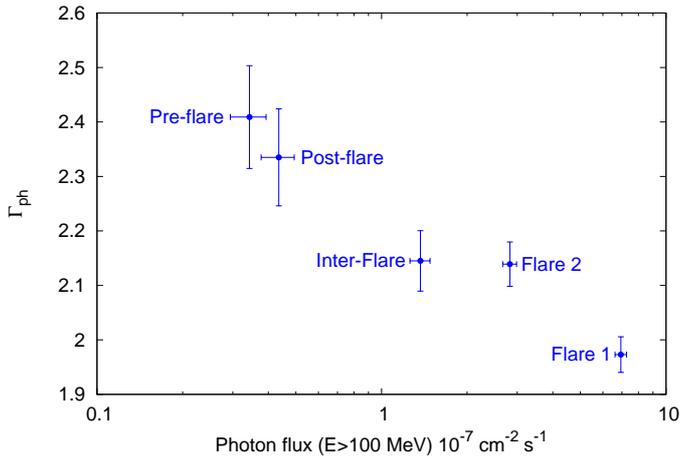}
 \caption{The $\gamma$-ray photon index, $\Gamma_\mathrm{ph}$, as a function of flux for the time periods
 defined in Table~\ref{table:timeintervals}.}
 \label{figure:fluxindex}
\end{figure}

Table~\ref{table:timeintervals} presents spectral analysis results for the
different $\gamma$-ray activity states of the source: ``pre-flare'' and
``post-flare'' periods represent the low-activity level, ``Flare\,1''
and~``Flare\,2''
represent the high-activity state, while during the ``interflare'' interval the
source showed an intermediate $\gamma$-ray flux level.
Figure~\ref{figure:gs} presents the observed {\em Fermi}/LAT spectrum during these states.
Significant evolution of the $\gamma$-ray photon index, $\Gamma_\mathrm{ph}$, is
detected between the different flux states (Table~\ref{table:timeintervals}).
Figure~\ref{figure:fluxindex} presents $\Gamma_\mathrm{ph}$ as a function of
($E>100$\,MeV) flux. The harder-when-brighter trend is clearly visible.

Integrating the \agile{} observations from 2009 November~18 12:00 (JD\,2455154.0) 
to 2009 November~22 12:00~UT (JD\,2455158.0), we obtain 
a \gray{} flux $F_{E>100\,{\rm MeV}} = (1.65 \pm 0.48)\times10^{-6}$\,\phcmsec{},
at a significance of $\sqrt{\mathrm{TS}}=6.1$.
This result is in good agreement with the flux value derived from the
preliminary analysis by \cite{Bulgarelli2009:ATel1310}.
Prior to the \textit{Fermi} launch, \agile{} observed \source{} (in pointing mode) 
during two other periods, but did not detect the source. 
During the first period
(from 2007 October~24 12:00~UT to 2007 November~1 12:00~UT, JD\,2454398.0--2454406.0),
the $2\sigma$ upper limit was $F_{E>100}\,{\rm MeV} \leq 0.28\times10^{-6}$\,\phcmsec{},
while in the second period (from 2008 April~30 12:00~UT to 2008 May~10 12:00~UT, JD\,2454587.0--2454597.0)
we obtained a $2\sigma$ upper limit of $F_{E>100\,{\rm MeV}} \leq
0.31\times10^{-6}$\,\phcmsec{}.

\subsection{X-ray to infrared spectrum}
\label{xraytoirspec}

Results of the X-ray spectral analysis are presented in Table~\ref{table:swiftlog}.
The obtained values of the X-ray photon index $\Gamma_\mathrm{ph~X-ray}$
are among the hardest reported for blazars \citep{2002babs.conf...63G,2005A&A...433.1163D,2009ApJ...704...38S}.
Radio-loud NLSy1 have $\Gamma_\mathrm{ph~X-ray}$ similar to the
ones found in blazars \citep{2013ApJ...768...52P,2009ApJ...707L.142A}. 
However, it cannot be excluded that the X-ray spectrum 
with an intrinsic value of $\Gamma_\mathrm{ph~X-ray}$ is artificially hardened 
by additional absorbing material 
along the line of sight (see the discussion of NOT imaging results below).
Future high-quality X-ray observations are                                                                                                        
necessary for investigating this possibility.

The UV and blue parts of the optical spectrum are flat, in 
contrast to the steep spectrum seen in the near-IR ($iJHK$ bands).
Also, the observed variability amplitude is decreasing toward bluer
wavelengths, with the $U$-band brightness being essentially constant. 

The $H$-band flux showed an increase of about            
one magnitude during the Flare\,2 period with respect to the Flare\,1 and
post-flare periods, in contrast to the behavior seen in other bands. 
The results of ground-based photometric measurements are summarized in
Table~\ref{table:photometry}.

\subsection{Imaging with NOT}
\label{sec:subsectionNOTimaging}

The Nordic Optical Telescope images (Fig.~\ref{figure:notr}) show 
a fuzzy extended object, probably a galaxy, with a point 
source offset $0\farcs6$ from its center. Considering that there are many galaxies 
of comparable brightness visible in the field, this picture may be
interpreted as the AGN (corresponding to the point source) shining
through an unrelated foreground galaxy. This may be the source of confusion
in the AGN's redshift determination
\citep{2009ATel.2306....1S,2008ApJS..175...97H,1998ApJ...494...47F}, and it
also explains the steepness of the optical-IR SED (the observed SED was
corrected for Milky Way absorption, but absorption in the
intervening galaxy may also be significant). On the other hand, it is
not uncommon for AGN host galaxies to have disturbed morphologies, making it
appear that the AGN is off-center.

The galaxy contributes a large fraction of the total optical flux, when the
point source is in the low state.
If the host galaxy of GB\,1310$+$487 is similar to the
giant ellipticals studied by \cite{2005ApJ...635..173S}, 
$\langle M_R \rangle = -22.9 \pm 0.5$\,mag, 
its magnitude at $z=0.500$ should be $R \approx 20$ 
(or $0.6$\,mag fainter at $z=0.638$; Sect.~\ref{sec:keckimgspec}). 
A typical NLSy1 from the \cite{2010A&A...518A..10V} catalogue having
$\langle M_V \rangle = -21.4$\,mag would appear 1\,mag fainter than a
giant elliptical in the $R$ band assuming $V-R=0.5$\,mag \citep{1996MNRAS.280....6X}.
Therefore, the observed
galaxy could be the host of GB\,1310$+$487. The visible offset between
the point source and the center of extended emission could result from the
disturbed morphology of the host, as noted above. 

\subsection{Keck imaging and spectroscopy}
\label{sec:keckimgspec}

\begin{figure}
 \centering
 \includegraphics[width=0.48\textwidth,trim=0.8cm 1.4cm 1.0cm 1.0cm,clip]{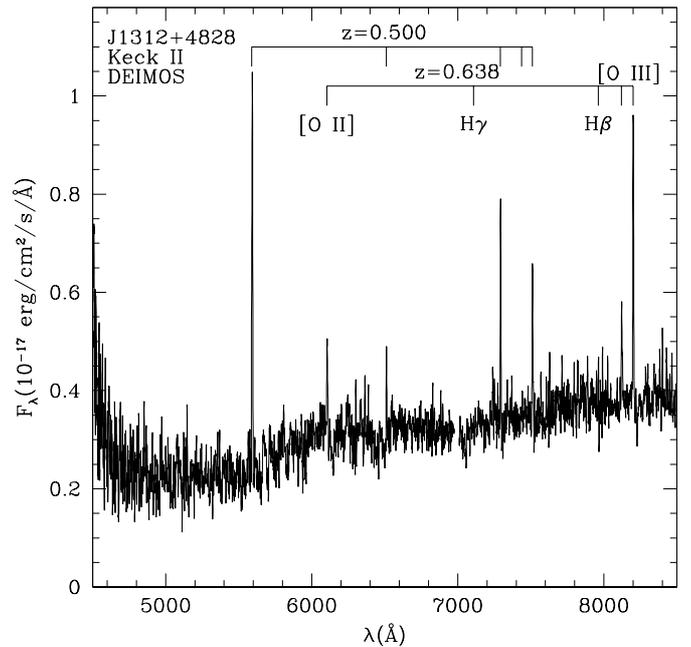}
 \caption{ Keck\,II/DEIMOS spectrum of 
GB\,1310$+$487. 
The two narrow-line systems are indicated. 
The region 6900--7000\,\AA\ is lost to a gap
between the two CCDs. The slit position is shown in
Figure~\ref{figure:keckimg}. }
 \label{figure:keckspec}
\end{figure}

Standard reductions, extractions, and calibrations of the DEIMOS data produced the
spectrum shown in Figure~\ref{figure:keckspec}; it is the
average of two observations conducted on April~07 and June~10, 2013. 
The strongest line, [O~II] $\lambda$3727, confirms the HET
redshift identification at $z=0.500$, and we also see [O~III] and narrow 
Balmer emission for this system. However, there are additional lines, mostly
in the red half. These represent a {\it second} system with narrow forbidden
and Balmer emission, this time at $z=0.638$. The line strengths are 
given in Table~\ref{table:EmLines}. 
The [O~II] doublets are barely resolved, but the oxygen
and Balmer line widths are consistent with the instrumental resolution.
The [O~III] emission at $z = 0.638$ appears resolved
with a deconvolved width of $\sim 200$\,km\,s$^{-1}$.
Unfortunately, the red
limit of the spectrum does not include the H$\alpha$/[N~II] lines for either
system. For the $z=0.500$ system, we cover [O~I]~$\lambda$6300, which is weak or
absent. For the $z=0.638$ system, we cover Mg~II $\lambda$2800, and can place a $3\sigma$ rest 
equivalent width limit of $\sim 1.0$\,\AA\ on any broad emission; 
the H$\beta$ line is marginally detected for the $z=0.638$ system at $4\sigma$
level. The ratio of H$\beta$ to [O~III] ($z=0.638$) is small,
even if the H$\beta$ flux is treated as an upper limit.
Together with the resolved [O~III] this indicates nuclear excitation.

%
%
\begin{table}
\caption{GB\,1310$+$487 emission-line strengths.}             
\label{table:EmLines}      
{
\centering                          
\begin{tabular}{@{}c@{\,\,}c@{\,\,}c@{\,}c@{\,\,}c@{\,}c@{}}        
\hline\hline                 
Species  & $\lambda_{\rm Rest}$ & Flux$_{0.500}^a$ & EW$_{0.500}^b$ & Flux$_{0.638}^a$ & EW$_{0.638}^b$ \\    
\hline                         
[O~II]     & 3726$+$3729        & $4.24\pm0.07$ & $32 \pm 6$ & $0.97\pm0.06$ & $4.2 \pm 0.5$  \\

H$\gamma$ &4340                & $0.65\pm0.05$ & $3.1 \pm 0.7$ & \multicolumn{2}{c}{cosmic-ray~hit}   \\

H$\beta$  &4861                & $1.69\pm0.06$ & $6.6 \pm 0.8$ & $0.24\pm0.06$ & $4.1 \pm 0.7$  \\

[O~III]    &4959                & \multicolumn{2}{c}{not~detected} & $1.00\pm0.06$ & $3.6 \pm 0.8$     \\

[O~III]    &5007                & $1.11\pm0.05$ & $6.5 \pm 0.6$ & $2.61\pm0.05$ & $12.0 \pm 2.2$ \\
\hline                                   
\end{tabular}
}

$^a$~Fluxes are in units of $10^{-17}$\,erg\,cm$^{-2}$\,s$^{-1}$, with $1\sigma$ 
statistical errors. The overall flux scale is uncertain
by up to a factor of 2, but the relative fluxes are much more accurate.
$^b$~The equivalent width (EW) in angstroms.
\end{table}

Thus, we clearly have two superimposed systems and wish to
identify which system hosts the radio-loud core (and, by inference, 
the $\gamma$-ray source). 
The Keck\,I LRIS images confirm
the basic structure seen in the NOT images; the source is extended with 
a brighter core displaced $\sim 0\farcs6$ to the west.
Figure~\ref{figure:keckimg}
shows $8\arcs$ regions around the AGN. The DEIMOS slit position on April~07
is marked on the $g$ frames (left). At the bottom we show the images after removal
of a point-source PSF ($g=23.89$, $R=22.45$ mag) from the offset core. 
The residuals show a relatively regular galaxy having FWHM =
$1\farcs8$, with $g=21.95$ and $R=20.59$ mag.  The coordinate 
system was referenced through the SDSS image of the field, with an
estimated uncertainty relative to the radio frame of $0\farcs2$;
the circles show the position of the VLBI source (Sect.~\ref{intro}) 
and have radii twice this uncertainty. Hence, the radio source is coincident
with the point-like peak of the combined source. We also find
that the $z=0.500$ emission lines are offset $0\farcs21 \pm 0\farcs06$ SE
along the slit from the $z=0.638$ system, 
toward the continuum tail representing the extended galaxy. The deprojected
offset is $\sim 0\farcs35$ E of the AGN core. We thus conclude
that the true AGN redshift is $z=0.638$, and we are viewing it through
an approximately face-on galaxy showing strong narrow-line emission.

Our extracted spectrum is weighted toward the AGN core, 
although it also contains appreciable light from the foreground galaxy. 
Both spectra are dominated by narrow forbidden lines, yet there is appreciable
continuum associated with both components as well. The foreground galaxy is 
probably not an AGN, but we cannot be certain; without the [N~II]/H$\alpha$ 
line ratio, we are unable to fully distinguish ``LINER'' (Low Ionization Nuclear Emission-line Region) emission from an 
H~II region \citep[e.g.,][]{1997ApJS..112..315H}. However, the strong [O~II] $\lambda$3727 and lack of obvious 
[O~I] $\lambda$6300 argue against a power-law ionizing 
spectrum, suggesting that the $z=0.500$ emission represents star formation 
in the foreground galaxy lacking AGN activity.

The positional accuracy of the available observations of multiwavelength
variability (Sect.~\ref{mwobse}) is not sufficient to distinguish between the foreground and background
objects discussed here and in Sect.~\ref{sec:subsectionNOTimaging} as the source 
of high-energy emission. The proposed
interpretation that the background AGN is the high-energy source rests on the
consideration that the observed fast $\gamma$-ray variability
(Sect.~\ref{sec:grayresults}) is typical of radio-loud AGNs (which the
background source is), while there are no firm indications of AGN activity
in the foreground galaxy.

\subsection{Results of radio observations}

\begin{figure}
 \centering
 \includegraphics[width=0.5\textwidth]{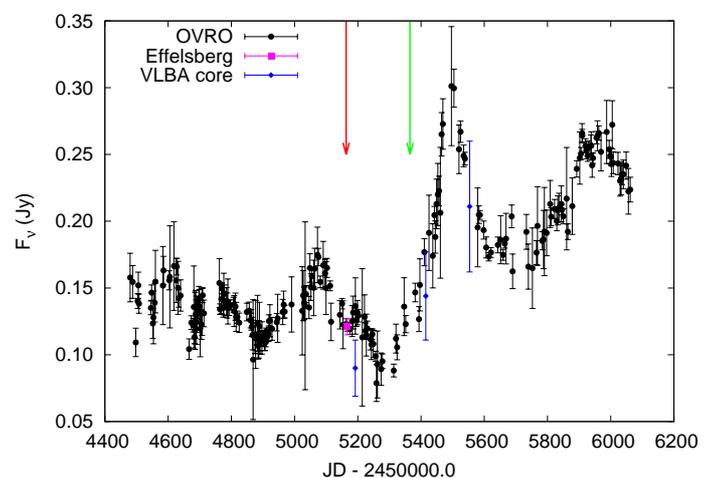}
 \caption{Radio lightcurve at 15\,GHz obtained with the OVRO 40\,m telescope
 (points) supplemented with two 14.6\,GHz measurements obtained with the
 Effelsberg~100\,m telescope (square). VLBA measurements of the
 core (component C0, Table~\ref{table:mojave}) are indicated as diamonds.
 The two arrows mark the peaks of
 the $\gamma$-ray flares observed by {\em Fermi}.}
 \label{figure:ovro}
\end{figure}

The radio spectrum of GB\,1310$+$487 is generally flat, with a wide peak
located between 22\,GHz and 86\,GHz 
(Table~\ref{table:mfreqradio}). 
The variability amplitude at 2.64\,GHz is slightly lower compared to higher
frequencies.
The 15\,GHz lightcurve of GB\,1310$+$487 obtained with the OVRO~40\,m
telescope and complemented by measurements with the Effelsberg~100\,m
and the VLBA is presented in Figure~\ref{figure:ovro}. 
It shows a period of high activity with two separate peaks that started in mid-2010
 and is still ongoing.

\begin{table}
\caption{Multifrequency radio observations of GB\,1310$+$487.}\label{table:mfreqradio}
{\centering
\begin{tabular}{rcc|rcc}
\hline  \hline
 ${\nu}$~~~ & $F_{\nu}$ & $\sigma$ & ${\nu}$~~~ & $F_{\nu}$ & $\sigma$ \\
 (GHz)  & (Jy) & (Jy) & (GHz)  & (Jy) & (Jy) \\
\hline
\multicolumn{3}{c}{RATAN-600~~~2003-06}         &\multicolumn{3}{c}{Effelsberg~100\,m~~~2010-06-28}\\
 21.74 &  0.401 &  0.104                          & 2.64  &   0.135  &   0.002                       \\
 11.11 &  0.222 &  0.012                          & 4.85  &   0.107  &   0.001                       \\
  7.69 &  0.157 &  0.019                          & 8.35  &   0.111  &   0.002                       \\
  3.95 &  0.221 &  0.057                          &10.45  &   0.112  &   0.005                       \\
\multicolumn{3}{c}{Effelsberg~100\,m~~~2009-12-01}&\multicolumn{3}{c}{Effelsberg~100\,m~~~2011-06-05}\\
 2.64  &   0.161  &   0.001                       & 2.64  &   0.201  &   0.006                       \\
 4.85  &   0.133  &   0.001                       & 4.85  &   0.189  &   0.003                       \\
 8.35  &   0.130  &   0.002                       & 8.35  &   0.207  &   0.005                       \\
10.45  &   0.130  &   0.003                       &10.45  &   0.213  &   0.010                       \\
14.60  &   0.121  &   0.006                       & & & \\
\multicolumn{3}{c}{IRAM~30\,m~~~2009-12-07}       & & & \\
 86.24 &    0.282  &   0.070                      & & & \\
142.33 &    0.206  &   0.065                      & & & \\
\hline
\end{tabular}
}

{\bf Column designation:}
Col.~1, the central observing frequency; and
Cols.~2,~3, the observed flux density and its uncertainty, respectively.
\end{table}

\begin{figure}
\centering
\includegraphics[width=0.48\textwidth,angle=270]{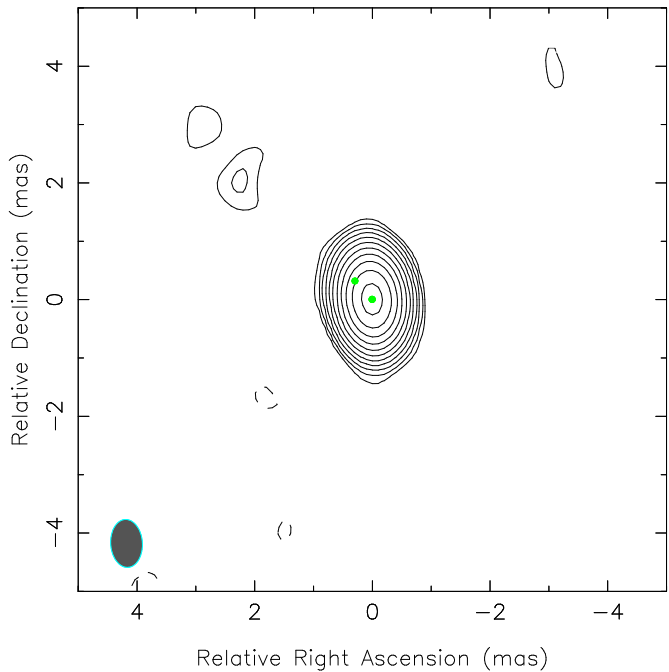}
\caption{VLBA radio image of GB\,1310$+$487 obtained on 2010-12-24 at 15\,GHz
during the course of the MOJAVE program. The image map peak is 0.206\,Jy beam$^{-1}$ and 
the first contour is 0.15\,mJy beam$^{-1}$. Adjacent contour levels are separated by a factor of 2. 
Naturally weighted beam size is indicated at the lower-left corner of
the image. Green circles indicate positions and best-fit sizes of 
the model components presented in Table~\ref{table:mojave}.}
\label{figure:vlbiimages}
\end{figure}

The 15\,GHz VLBA images (Fig.~\ref{figure:vlbiimages}) show two emission regions 
separated by $\sim 0.4$\,mas. To quantify
their parameters we fit the observed visibilities with 
a model consisting of two circular Gaussian components using the
\texttt{Difmap} software \citep{1997ASPC..125...77S}. The
modeling results are presented in Table~\ref{table:mojave}.
The uncertainties in parameters of the model components were estimated following
\cite{2008AJ....136..159L}, and the resolution limit achieved for each
component was computed following
\cite{2005astro.ph..3225L} and \cite{2005AJ....130.2473K}.

The SW component increased its brightness during
the three epochs. Comparison with the lightcurve in Figure~\ref{figure:ovro}
shows that this component is responsible for most of the flux
observed with single-dish instruments. The fainter component located to the 
NE is gradually fading. If the SW component is the
15\,GHz core, the position of the second component aligns nicely with the
orientation of the kiloparsec-scale jet observed with the VLA at 1.4\,GHz by
\cite{1983AJ.....88.1591M}. No significant proper motion could be
detected between the three 15\,GHz MOJAVE epochs. The $3\sigma$ upper limit 
which can be placed on proper motion is $\mu < 0.3$\,mas\,yr$^{-1}$, 
corresponding to $\beta_\mathrm{app} < 11$ 
($\beta_\mathrm{app}$ is in units of the speed of light) at
the source redshift, which is within the range of apparent jet
speeds occupied by $\gamma$-ray-bright blazars \citep{2009ApJ...696L..22L,2010A&A...512A..24S}.
The projected linear size of the double structure resolved with the VLBA is
$\sim 2.7$\,pc $=8 \times 10^{18}$\,cm.
The overall 15\,GHz VLBI polarization of the source measured by MOJAVE
is 2.8--6.5\% which is indicative of beamed blazar emission. 
The weakly beamed, high viewing angle sources in MOJAVE tend to 
be unpolarized \citep{2005AJ....130.1389L}.

\begin{table}
\caption{Parsec-scale components observed at 15\,GHz.}
\label{table:mojave}      
{\centering          
\begin{tabular}{c c c c c c c}   
\hline\hline       
Comp. &    Distance      &   FWHM &  Flux density  &   $T_b$     \\
          &     (mas)    &  (mas) &  (Jy)  &   (K)       \\
\hline                    
\multicolumn{5}{c}{2009-12-26 = JD\,2455192}\\
C0 &  $\dots$           & $<0.39$ &  $0.090 \pm 0.021$ & $>3\times10^{9}$ \\
C1 &  $0.32 \pm 0.11$ & $<0.42$ &  $0.014 \pm 0.004$ & $>4\times10^{8}$ \\
\multicolumn{5}{c}{2010-08-06 = JD\,2455415}\\
C0 &  $\dots$           & $<0.36$ &  $0.144 \pm 0.033$ & $>6\times10^{9}$ \\
C1 &  $0.36 \pm 0.10$ & $<0.40$ &  $0.008 \pm 0.002$ & $>3\times10^{8}$ \\
\multicolumn{5}{c}{2010-12-23 = JD\,2455554}\\
C0 &  $\dots$           & $<0.40$ &  $0.211 \pm 0.049$ & $>7\times10^{9}$ \\
C1 &  $0.43 \pm 0.12$ & $<0.52$ &  $0.004 \pm 0.002$ & $>8\times10^{7}$ \\
\hline                  
\end{tabular}
}

{\bf Column designation:}
Col.~1, component name, where C0 is the presumed core and C1 is the
decaying jet component;
Col.~2, projected distance from the core (C0);
Col.~3, FWHM of the Gaussian component;
Col.~4, component flux density; and
Col.~5, observed brightness temperature.

\end{table}

\subsection{SED during the two flares}

\begin{figure*}
 \centering
 \includegraphics[height=0.78\textwidth,angle=270,trim=0.0cm 0.0cm 0.05cm 0.0cm,clip]{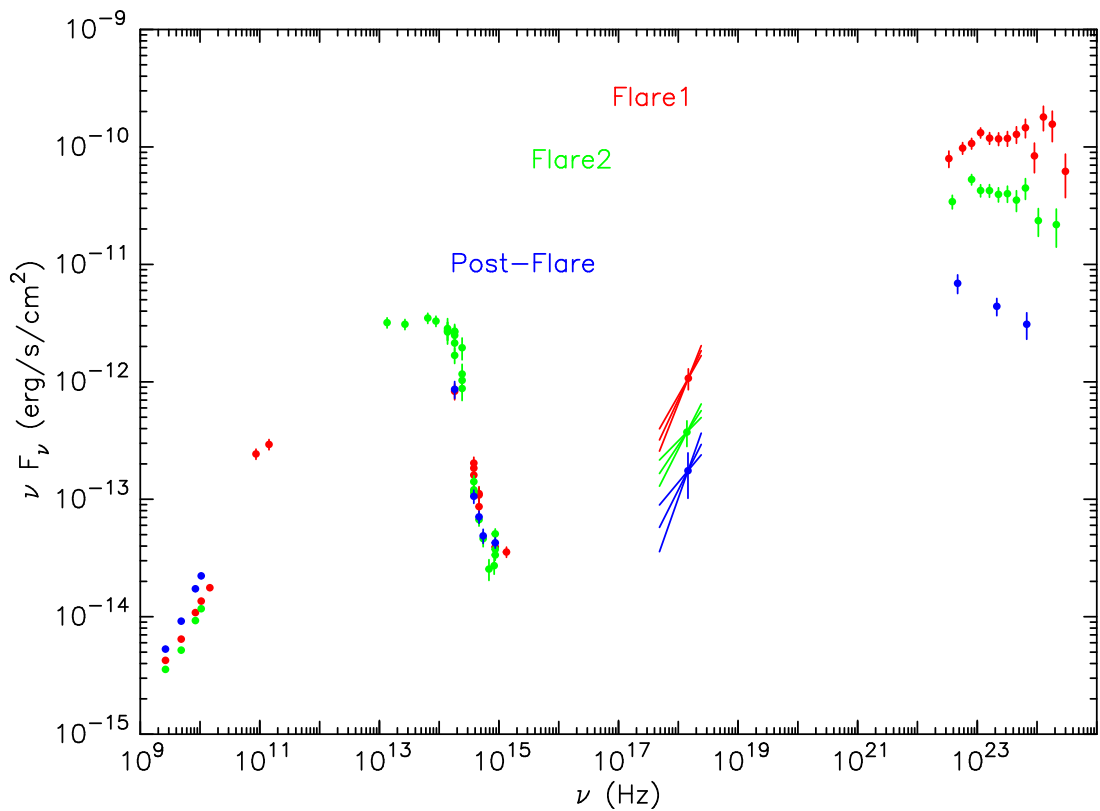}
 \caption{Quasi-simultaneous radio to $\gamma$-ray SED of GB\,1310$+$487 during the two
 flaring episodes and the post-flare period covered by our multiwavelength
 observations. The time intervals corresponding to these events are defined
 in Table~\ref{table:timeintervals}.}
 \label{figure:sed}
\end{figure*}

The SED of GB\,1310$+$487 is presented in
Figure~\ref{figure:sed}. It has the classical two-humped shape with the
high-energy hump dominating over the synchrotron hump during the first
(brighter) flare by a Compton dominance factor of $q \ge 10$.
For the Flare\,2 period the value of Compton dominance may be 
measured accurately thanks to simultaneous observations of {\em Fermi}/LAT and WISE:
$q=12$. The uncertainty of this measurement is limited by the accuracy of 
the absolute calibration of the two instruments and should be less
than 10\%. Fast variability within the Flare\,2 period may also contribute to the
uncertainty. The value of $q$ is probably larger for Flare\,1 than for Flare\,2, judging 
from the lower near-IR flux observed during Flare\,1.

\section{Discussion}
\label{discussion}

\subsection{$\gamma$-ray luminosity, variability, and spectrum}
\label{discussiongamma}

The monochromatic $\gamma$-ray energy flux averaged over the duration of the first flare
is $\nu F_\nu \approx 10^{-10}$\,erg\,cm$^{-2}$\,s$^{-1}$. At the redshift of the
source this corresponds to an isotropic luminosity of 
$\sim 10^{47}$\,erg\,s$^{-1}$.
Considering the expected bolometric correction of a factor of a few, 
the $\gamma$-ray luminosity of GB\,1310$+$487 is comparable to that
typically observed in flaring $\gamma$-ray blazars
\citep[e.g.,][]{2011ApJ...733...19T,2010ApJ...721.1425A,2010ApJ...722..520A}
and NLSy1 \citep{2013arXiv1303.3030D}. It is about two orders of magnitude lower 
than the most extreme GeV flares of 3C\,454.3 in November~2010
\citep{2011ApJ...733L..26A} and PKS\,1622--297 in June~1995
\citep{1997ApJ...476..692M}. The outstanding $\gamma$-ray flare of
3C\,120 in November~1968 had a comparable isotropic luminosity of $\sim 10^{47}$\,erg\,s$^{-1}$
\citep{1972SvA....15..879V}. The exceptional GeV photon flux of 3C\,120 was 
due to the relative proximity of the source ($z=0.033$;
\citealt{1988PASP..100.1423M}) compared to the brightest $\gamma$-ray
blazars mentioned above.
The observed large $\gamma$-ray luminosity of GB\,1310$+$487 is an indirect indication of 
a high Doppler boosting factor of the source \citep{2007ApJ...671.1355T,2009A&A...507L..33P}.

The difference in lightcurve shape, overall duration, and shortest 
observed variability timescale between the two flares of the source 
may indicate that they occurred in different jet 
regions or were powered by different emission mechanisms as discussed below.
In both cases, this implies differences in the emitting-plasma parameters 
for the two flares, such as the electron energy distribution, 
magnetic field strength, bulk Lorentz factor, or external photon field strength.
Variability timescales of $3$~days and shorter are common in GeV blazars
\citep[e.g.,][]{1997ApJ...476..692M,2010ApJ...710..810A,2011ApJ...733L..26A,2011AdSpR..48..998S}.
The light-travel-time argument 
limits the $\gamma$-ray emitting region size $r < c \delta t_\mathrm{var} / (1+z)
\approx \mathrm{a~few} \times 10^{15}$\,cm, where $c$ is the
speed of light in vacuum, $z$ is the source redshift, and
conservatively assuming the Doppler factor 
$\delta \equiv [\Gamma (1-\beta \cos \theta)]^{-1}<\mathrm{a~few}$ 
(we have no evidence of extreme Doppler boosting from VLBI and
$\gamma$-ray data; Sect.~\ref{dopplerfactor}), where
$\Gamma$~is the Lorentz factor, $\beta$~is the bulk velocity of the emitting
blob in units of the speed of light, and $\theta$~is the angle between the
blob velocity and the line of sight.

It is important to check that the observed harder-when-brighter 
trend in the $\gamma$-ray spectrum is not related to the
expected correlation between the flux and the index in the power--law
model. If the number density of photons arriving from the source is
$\mathrm{d}N/\mathrm{d}E = N_0 (E/E_0)^{-\Gamma_\mathrm{ph}}$
(where $N_0$, and $E_0$ are constants), the integrated photon flux between
energies $E_\mathrm{max}$ and $E_\mathrm{min}$ is
$$F = \frac{N_0 E_0}{-\Gamma_\mathrm{ph}+1} \left[
\left(\frac{E_\mathrm{max}}{E_0}\right)^{-\Gamma_\mathrm{ph}+1} -
\left(\frac{E_\mathrm{min}}{E_0}\right)^{-\Gamma_\mathrm{ph}+1} \right].$$
Assuming that $\Gamma>1$ and $E_\mathrm{max}$ is large,
$$F \approx -\frac{N_0 E_0}{-\Gamma_\mathrm{ph}+1}
\left(\frac{E_\mathrm{min}}{E_0}\right)^{-\Gamma_\mathrm{ph}+1}.$$
The derivative 
$$\frac{\mathrm{d}F}{\mathrm{d}\Gamma_\mathrm{ph}} = F \left[
\frac{1}{-\Gamma_\mathrm{ph}+1} -
\ln\left(\frac{E_\mathrm{min}}{E_0}\right) \right] \neq 0$$
if $\ln(E_\mathrm{min}/E_0)\neq1/(-\Gamma_\mathrm{ph}+1)$.
The range of parameters derived from our analysis is 
$\Gamma =1.97$--$2.41$, $E_\mathrm{min}=100$\,MeV, $E_0=283$\,MeV, and
$-1.04=\ln(E_\mathrm{min}/E_0)<1/(-\Gamma_\mathrm{ph}+1)=-1.03$ to $-0.70$,
so $\mathrm{d}F/\mathrm{d}\Gamma_\mathrm{ph}>0$. The expected correlation
between $\mathrm{d}F$ and $\Gamma_\mathrm{ph}$ due to their mathematical dependence 
is positive, which is opposite to what is actually observed. We conclude
that the observed harder-when-brighter trend is real and not related to
the intrinsic correlation of the model parameters.

Previously a harder-when-brighter trend (Fig.~\ref{figure:fluxindex}) 
has been seen at GeV energies only in a handful of blazars:
3C\,273, PKS\,1502$+$106, AO\,0235$+$164, and 4C\,+21.35 by {\em Fermi}
\citep{2010ApJ...714L..73A,2010ApJ...710..810A,2010ApJ...710.1271A,2011ApJ...733...19T};
3C\,454.3
\citep{2010ApJ...721.1383A,2010ApJ...712..405V,2011ApJ...733L..26A,2011MNRAS.417L..11S}
and PKS\,1510$-$089 by {\em Fermi} and \agile{}
\citep{2010ApJ...721.1425A,2011A&A...529A.145D}; and
3C\,279 \citep{2001ApJ...553..683H} and PKS\,0528$+$134 \citep{1996ApJ...470..831M} by
EGRET. The same harder-when-brighter behavior was suggested by the combined analysis of
relative spectral index change as a function of relative flux change in a
few of the brightest FSRQs, low-, and intermediate-peaked BL~Lacs using the first
six~months of {\em Fermi} data by \cite{2010ApJ...710.1271A}.
The current detection presents one of the clearest examples of this spectral behavior.

The spectral evolution during the Flare\,1--interflare--Flare\,2 periods
(Fig.~\ref{figure:gs} and Table~\ref{table:timeintervals}) may be
qualitatively understood as the gradual decrease in energy of 
the $\gamma$-ray emission peak. During Flare\,1, the spectrum is hard,
implying that the spectral peak is located above or around 5\,GeV. Consequently, during
the interflare period the spectrum is softer, with a hint of curvature;
the emission peak may be located around 1--2\,GeV. 
Later, during Flare\,2, the spectrum is also soft with a peak
possibly located at even lower energies. This interpretation is inspired by
the visual inspection of Figure~\ref{figure:gs}. 
The peak-frequency evolution is difficult to quantify owing to the insufficient 
number of collected photons, which results in the simple PL fit (with no curvature) 
being a statistically acceptable model for the LAT data. However, the overall SED
(Fig.~\ref{figure:sed}) suggests that the high-energy emission peak
should be located somewhere around the LAT band. We can use the LAT
spectral index ($\Gamma_\mathrm{ph}$) vs. Compton peak frequency 
($\nu_{\rm peak~Hz}^{\rm IC}$) correlation
$\log_{10} \nu_{\rm peak~Hz}^{\rm IC} = -4.0\Gamma_\mathrm{ph} +31.6$ reported by
\cite{2010ApJ...716...30A} to estimate that 
$\nu_{\rm peak~Hz}^{\rm IC}$ changed from $10^{22}$ to $10^{24}$\,Hz
between the pre-flare and Flare\,1 periods.
The change in $\nu_{\rm peak~Hz}^{\rm IC}$ may result not from a
continuous shift of a single $\gamma$-ray emission peak, but from 
a change in relative strengths of two emission components
peaking at different frequencies, 
as discussed in Sect.~\ref{interpretation}.

\subsection{Jet Doppler factor}
\label{dopplerfactor}

The Doppler factor of the relativistic jet in GB\,1310$+$487 may be constrained
using two independent lines of argument: one based on the
requirement that the emitting region should be transparent to its own
$\gamma$ radiation (since we observe it), the other based on the absence of
apparent proper motion seen by the VLBA.

The minimum Doppler factor needed to avoid $\gamma-\gamma$ 
attenuation for $\gamma$-rays interacting with lower energy photons present 
inside the emitting region may be calculated using Eq.~(39) of
\cite{2008ApJ...686..181F},
$$
\delta_{\gamma\gamma} > \left[ \frac{ 2^{a-1} (1+z)^{2-2a} \sigma_T D_L^2}
          {m_e c^4 t_\mathrm{var} } \epsilon_1 f_{\epsilon_1^{-1}}^{syn}
      \right]^{\frac{1}{6-2a}},
$$
where it is assumed that the synchrotron flux is well represented by a
power law of index $a$ ($f^{\rm syn}_{\epsilon} \propto \epsilon^a$),
$\sigma_T$ is the scattering Thomson cross-section,
$D_L$ is the luminosity distance to the source,
$m_e$ is the electron mass, and
$\epsilon_1 = E / ( m_e c^2 )$ is the dimensionless energy of a $\gamma$-ray
photon with energy $E$ for which the optical depth of the emitting
region $\tau_{\gamma\gamma}=1$ (see also \citealt{1995MNRAS.273..583D}).
The maximum energy of observed $\gamma$-ray photons that can be attributed
to the source is $\sim 10$\,GeV (Fig.~\ref{figure:gs}), so $\epsilon_1 = 
10\mathrm{\,GeV} / (5.11 \times 10^{-4})\mathrm{\,GeV} = 2\times 10^{4}$.
This means $\epsilon_1^{-1} = 5.11 \times 10^{-5}$, and the corresponding frequency for this
is $6.3\times 10^{15}$\,Hz. From the observed SED (Fig.~\ref{figure:sed}), we estimate  
$f_{\epsilon_1^{-1}}^{\rm syn} \approx 10^{-14}\mathrm{\,erg\,s}^{-1}\mathrm{\,cm}^{-2}$ and $a \approx -2$. 
Taking $t_\mathrm{var} = 3$\,days (Sect.~\ref{sec:grayresults})
we obtain $\delta_{\gamma\gamma} > 1.5$.

Assuming the angle between the jet axis and the line of sight, $\theta$, is
$\theta_\mathrm{max}$, the one that maximizes the apparent speed, 
$\beta_\mathrm{app}$, for a given intrinsic
velocity, $\beta$, we may estimate the corresponding Doppler factor
$\delta_\mathrm{VLBA} < 11$ (if
$\delta_\mathrm{VLBA} = \Gamma = \sqrt{\beta_\mathrm{app}^2+1}$).
We note that if $\theta$ is smaller than $\theta_\mathrm{max}$, 
the actual $\delta$ will be larger than the above estimate 
(see, e.g., \citealt{2007ApJ...658..232C,2007Ap&SS.311..231K,2009arXiv0909.2576M} for a
discussion of relativistic kinematics in application to VLBI).

Recent RadioAstron \citep{2013ARep...57..153K} Space--VLBI observations of high brightness 
temperatures in AGNs suggest that the actual jet flow speed is often higher than the 
jet pattern speed \citep{2013arXiv1303.5451S}. These results question the applicability
of $\delta$ estimates based on VLBI kinematics. The available lower limits on
the core brightness temperature, $T_b$, in GB\,1310$+$487  (Table~\ref{table:mojave}) are
consistent with negligible Doppler boosting within the standard assumption
of the equipartition inverse-Compton limited $T_b \approx \mathrm{a~few}\times10^{11}$\,K
\citep{1994ApJ...426...51R}.

\subsection{Black hole mass}

If we equate the linear size estimated from the shortest observed variability 
timescale to the Schwarzschild radius, 
the corresponding black~hole mass would be 
$M_\bullet \approx 10^{10}\, {\rm M}_\odot$. However, TeV observations of ultra-fast (timescale of minutes)
variability in blazars PKS\,2155$-$304
\citep{2007ApJ...664L..71A,2010A&A...520A..83H} and
Mrk\,501 \citep{2007ApJ...669..862A} 
lead to $M_\bullet$ estimates inconsistent
with those obtained by other methods \citep{2008MNRAS.384L..19B}, unless
an extremely large Doppler factor
$\delta \approx 100$ is assumed for the $\gamma$-ray emitting region
\citep{2008MNRAS.386L..28G,2011AdSpR..48..998S}. 
Short timescale variability may arise from the interaction 
of small (size $r<r_s$) objects such as stars \citep{2012ApJ...749..119B} or 
BLR clouds \citep{2010A&A...522A..97A} with a broad relativistic jet.
This should caution us against putting much trust in the above $M_\bullet$ estimate.

\subsection{UV, optical, and IR emission}

The UV-to-IR behavior of the source may be understood if the near-IR light is
dominated by the synchrotron radiation of the relativistic jet, while in the
optical--UV the contribution of line emission 
and/or thermal emission from the accretion disk starts to dominate over the synchrotron radiation. 
The line and thermal emission are not relativistically beamed and, therefore, more stable compared
to the beamed synchrotron jet emission, decreasing the variability amplitude in the
parts of the SED where their contribution to the total light is comparable
to that from the jet. 
Thermal emission features are observed in SEDs of many FSRQ-type blazars
\citep[e.g.,][]{2006A&A...453..817V,2009ARep...53..510H,2010ApJ...721.1425A,2011A&A...529A.145D}.
Starlight from the host galaxy also contributes to the total optical
flux in some $\gamma$-ray loud AGN
\citep{2007A&A...475..199N,2011ApJ...727..129A,2011ApJ...736..131A}.
This contribution is significant mostly for BL~Lac-type blazars and
non-blazar AGN.
Finally, as discussed above, the nearby galaxy and the star-like object, both
probably unrelated to the source under investigation,
may contribute to its total optical flux if an observation lacks angular 
resolution to separate contributions from these objects.

\subsection{Radio properties}
\label{sec:radioproperties}

The radio loudness parameter, $R_{\rm radio}$, defined as the ratio of 5\,GHz flux
density, $L_{5\mathrm{\,GHz}}$, to the $B$-band optical flux density, $L_B$, is 
$R_{\rm radio} =L_{5\mathrm{\,GHz}}/L_B \approx 10^4$. This is an order of magnitude larger than
typical $R_{\rm radio}$ values found in quasars \citep{1989AJ.....98.1195K} 
and radio-loud NLSy1 galaxies \citep{2006PASJ...58..829D}, but it is comparable
to the largest observed values
\citep{2013ApJ...764...43S}\footnote{\cite{2013ApJ...764...43S} use the rest
frame luminosity at 5\,GHz and $2500\,\AA$ respectively to define $R$. 
This definition of R should be consistent within a factor of a few with
the one we use.}. The extremely low optical luminosity compared to the
radio luminosity may either be an intrinsic property of this source, or it may
result from absorption in the intervening galaxy
(Sect.~\ref{sec:subsectionNOTimaging}, \ref{sec:keckimgspec}).

The radio spectrum of GB\,1310$+$487 (Table~\ref{table:mfreqradio}) 
is typical for a blazar. Relatively rapid (timescale of months) and 
coherent changes across the cm band suggest
that most of the observed radio emission comes from a compact region 
no more than a few parsecs in size.
Comparison of the 15\,GHz lightcurve presented in Figure~\ref{figure:ovro}
with the 15\,GHz VLBA results (Table~\ref{table:mojave}) indicates that 
the component C0 (presumably the core) is the one responsible for most
of the observed single-dish flux density of the source. Specifically, C0 is the site
of the major radio flare peaking around JD\,2455500 (October--November 2010).

The presence of correlation between cm-band radio and $\gamma$-ray emission 
is firmly established for large samples of blazars
\citep{2011ApJ...741...30A,2012A&A...537A..32A,2012ApJ...744..177L,2009ApJ...707L..56K}.
The typical $\gamma$-ray/radio time delay ranges 
from 1~month to 8~months in the observer's frame, with $\gamma$-rays leading radio
emission \citep{2010ApJ...722L...7P,2011A&A...532A.146L}. However, for individual sources it is
often difficult to establish a statistically significant correlation because
of the limited time span of simultaneous $\gamma$-ray--radio data compared to
a typical duration of radio flares \citep{2012arXiv1205.0276M}. This could also limit our knowledge of the maximum
possible radio/$\gamma$-ray time delay.

In the case of GB\,1310$+$487, no clear connection is visible between its radio 
and $\gamma$-ray activity, based both on the available 
single-dish (Fig.~\ref{figure:ovro}) and VLBI monitoring data
(Table~\ref{table:mojave}).

\subsection{Object classification}
\label{sec:classificationdiscussion}

\cite{2012ApJ...748...49S} classified the optical spectrum of GB\,1310$+$487 as a LINER, which is inconsistent with the $\gamma$-ray and radio loudness
(Sect.~\ref{sec:radioproperties}; \citealt{1988A&A...203...39G}). The absence of broad lines
precludes classification as a quasar. Prominent forbidden emission
lines are not typical of BL~Lac-type objects. Therefore, while being similar
to blazars in its high-energy, radio, and IR properties, GB\,1310$+$487
cannot be classified as a classical blazar on the basis of its optical
spectrum.

As discussed in Sect.~\ref{sec:keckimgspec}, 
the point-source emission at $z=0.638$ observed by Keck 
is most likely related to the AGN.
\cite{2000NewAR..44..381P} defines NLSy1 as having permitted lines only slightly
stronger than forbidden lines, [O~III]/H$\beta <3$, and FWHM(H$\beta$) $<
2000$\,km\,s$^{-1}$. The anomalously strong [O~III] $\lambda\lambda$4959, 5007 emission formally disqualifies
this source, and tends to support a Seyfert~2 
(or a narrow-line radio galaxy, considering the object's radio loudness)
classification, which would
be difficult to understand if the radio- and $\gamma$-ray jets align with
the Earth's line of sight.  Our S/N is too low to allow unambiguous
detection of Fe~II emission. Broad H$\beta$, if present, is weaker 
than a third of the narrow component, and there is no evidence of 
Mg~II~2800\,\AA. Thus, no broad-line component is observed. We also find
that Ca~H\&K are weak, if present, and the 4000\,\AA\ break is smaller than 
0.1. These aspects suggest appreciable nonthermal luminosity for the core
AGN emission. Thus, we tentatively advance the view that synchrotron
emission from the AGN dominates the variable point-source core, but that
a surrounding narrow line region dominates the line flux.

\cite{2014MNRAS.438.1149G} studied the WISE infrared colors of radio-loud AGN;
GB\,1310$+$487 falls in a region of the color diagram occupied mostly by
quasars and broad-line radio galaxies, although some narrow-line radio
galaxies are also present. The source GB\,1310$+$487 is well away from the locus of
low-excitation radio galaxies (LERGs) and also has a 22\,$\mu$m luminosity 
($\sim 4 \times 10^{45}$\,\ergsec) typical of high-excitation radio galaxies
(HERGs; see Fig.~8 in \citealt{2014MNRAS.438.1149G}). However, if one uses the
criteria of \cite{1997MNRAS.286..241J} GB\,1310$+$487 would qualify as a LERG
based on its optical spectrum. We note that WISE photometry of the AGN might
be contaminated by the foreground galaxy.

Being a narrow-line radio-loud AGN, the object is not a
member of common types of $\gamma$-ray flaring extragalactic
sources (blazars and NLSy1s). One possibility is that the object is
analogous to nearby radio galaxies like Per~A 
with additional amplification 
due to gravitational lensing that
makes $\gamma$-ray emission from its core detectable
at high redshift. 
The similarity to Per~A is supported by 
its lack of superluminal motion \citep{2013AJ....146..120L}, 
low $\delta$ inferred from SED modeling \citep{2010ApJ...719.1433A}, and
absence of changes in VLBI and single-dish radio properties that can be
attributed to GeV events \citep{2012MNRAS.423L.122N}.

Another possibility is that the object may be a bona fide
blazar with its optical non-thermal emission swamped by the host elliptical
as proposed by \cite{2013MNRAS.431.1914G} as possible counterparts of
unassociated {\em Fermi} sources. In this case, however, one would not
understand the observed variable optical point source. Higher S/N
spectroscopy with increased wavelength coverage would 
be helpful in characterizing the $z=0.638$ $\gamma$-ray/radio AGN.
Higher resolution spatial imaging is needed to probe the nature of 
the foreground ($z=0.500$) galaxy.

\subsection{Gravitational lensing}
\label{sec:gravlens}

Considering that the AGN is located behind the visible
disk of another galaxy (Sect.~\ref{sec:keckimgspec},
\ref{sec:classificationdiscussion}), amplification of the AGN light by
gravitational lensing is a real possibility. 
In the simplified case of a point lens, the AGN light is amplified by a
factor of
$A=(u^2+2)/(u\sqrt{u^2+4})$
\citep{1986ApJ...304....1P,1991ApJ...366..412G,2006astro.ph..4278W}, where $u$ is the ratio of the AGN/lensing-galaxy
separation ($0\farcs6$) to the lensing galaxy's Einstein radius,
$$R_E = \sqrt{ \frac{4GM_\mathrm{lens}}{c^2} \frac{D_\mathrm{lens-to-AGN}}{D_\mathrm{lens} D_\mathrm{AGN}}} \approx 1\farcs1
(M_{\mathrm{lens}~12})^{1/2},$$
where $G$~is the gravitational constant, $M$ is the lensing galaxy mass,
$D_\mathrm{lens}=1300$\,Mpc is the angular size distance to the lens,
$D_\mathrm{AGN}=D_A=1400$\,Mpc is the angular size distance to the AGN,
$D_\mathrm{lens-to-AGN}=D_\mathrm{AGN} - (1+z_\mathrm{lens})/(1+z_\mathrm{AGN}) D_\mathrm{lens}$
is the distance between the lens located at redshift $z_\mathrm{lens}$
and the AGN located at redshift $z_\mathrm{AGN}$, and
$M_{\mathrm{lens}~12}$ is $M$ expressed in the units of $10^{12}\,{\rm M}_\odot$.
The above amplification factor estimate involves a number of
simplifications including {\it (i)} the simplified lens geometry,
{\it (ii)} use of the observed AGN--lens separation which is larger than
the true one, and
{\it (iii)} the \cite{1986ApJ...304....1P} formula referring to the combined
light of two images (we know from observations that the single observed
image of GB\,1310$+$487 is much brighter than its second undetected image).
Taking into account these caveats, we estimate that the AGN image is
probably amplified by a factor of a few.

Since no second image is visible in optical Keck and
radio VLBA (this work) and VLA (JVAS survey; \citealt{1999MNRAS.307..225K}) data, 
we assume that its contribution to the source lightcurve at
other bands, including GeV, is also negligible. The absence of an observable
second image may indicate that either the lens is not massive 
and the source is still outside
its $R_E$ or the mass distribution in the lens is asymmetric and the lensed source
is close to a fold or cusp caustic.
If the lens is a singular isothermal sphere (SIS; e.g.,
\citealt{1994RPPh...57..117R,2006glsw.conf.....M}) and the lensed AGN is just outside
the Einstein radius 
defined for an SIS through the lensing galaxy's velocity dispersion
$\sigma_\mathrm{SIS}$,
$$R_{E\mathrm{\,SIS}} = 4 \pi \frac{\sigma_\mathrm{SIS}^2}{c^2}
\frac{D_\mathrm{AGN}}{D_\mathrm{lens}},$$
then a single image is formed having, in principle, an arbitrarily 
large amplification factor 
$A = 1/(1 - 1/u)$
\citep[e.g.,][]{1994A&A...286..748W}. Taking $R_{E\mathrm{\,SIS}} \le
0\farcs6$ we estimate $\sigma_\mathrm{SIS} \le 140$\,km\,s$^{-1}$ and the 
mass inside $R_{E\mathrm{\,SIS}}$ of $\le 6 \times 10^{10}\,{\rm M}_\odot$
\citep{1994A&ARv...5..239F}.
The low lensing galaxy mass, necessary to put the AGN image outside $R_E$
(and form a single image),
may be reconciled with its brightness if the galaxy is undergoing intensive
star formation, as indicated by strong emission lines in its spectrum.

In general, gravitational lenses producing a single magnified image of a
distant source should be more common than lenses producing multiple images.
However, most gravitational lens searches (like the JVAS-CLASS survey;
\citealt{2003MNRAS.341...13B}) are designed to identify only 
multiple-image lenses. The BL~Lac-type object AO\,0235$+$164 is an example
of a blazar shining through an intervening galaxy and having a single image 
weakly amplified by macrolensing \citep{1993ApJ...415..101A}.

A possibility of microlensing by individual 
foreground stars in the $z=0.5$ galaxy cannot be excluded. 
The timescale of such microlensing events may be estimated as a ratio of the
Einstein radius for a single star to the proper motion of the lens and is on the order of
tens of years. Therefore, microlensing is probably unrelated to the observed fast high-energy
variability, but may provide a significant
amplification that is nearly constant over the duration of our observations.
A large constant amplification due to microlensing could also explain the
absence of the second lensed image.

\subsection{Emission model constraints from the SED}

The high-energy SED hump dominates over the synchrotron hump during the first
(brighter) flare by a factor of $q > 10$. This is commonly observed in FSRQ-type blazars.
In the framework of the leptonic model, the large $q$ suggests 
that most of the observed $\gamma$-ray flux during the first flare 
should be attributed to the EC process rather than to the SSC scenario. 
An SSC model would require a large deviation from equipartition: 
the emitting-particle energy density would need to be $q^2$ times greater 
than the magnetic field energy density in the source frame 
\citep{2009ApJ...704...38S}, which is not expected; EC models do not require
deviation from equipartition to explain large values of $q$.

The dramatic difference in spectral slopes in the radio and IR
regions suggests the presence of a break in the electron energy distribution.
At the post-flare state, GB\,1310$+$487 shows a steep $\gamma$-ray
spectrum. If this spectrum is dominated by the SSC emission
(Sect.~\ref{interpretation}), it is possible
to estimate the Lorentz factor, $\gamma_\mathrm{b}$, of electrons emitting 
at the energy distribution break from the positions of the SSC
($\nu_\mathrm{SSC} \approx 10^{22}$\,Hz) and synchrotron ($\nu_\mathrm{syn}
\approx 10^{14}$\,Hz) SED peaks \citep{2009ApJ...704...38S}:
$\gamma_\mathrm{b} \approx (\nu_\mathrm{SSC}/\nu_\mathrm{syn})^{1/2} \approx 10^4$. 
This value is an order of magnitude larger than those found in detailed SED 
modeling of FSRQs,
\citep[e.g.,][]{2011ApJ...736L..38V,2012ApJ...754..114H,2013ApJ...779..174D}. 
Different SED models applied to the same source may lead to different 
estimates of $\gamma_\mathrm{b}$ \citep{2010arXiv1006.3084S,2010ApJ...721.1425A}.

This is also the case for GB\,1310$+$487 (Table~\ref{table:swiftlog}).
Hadronic models predict that X-rays are produced by synchrotron 
radiation of the secondary ultra-relativistic population of electrons and 
positrons. To reproduce the hard X-ray spectra observed in GB\,1310$+$487
(Table~\ref{table:swiftlog}) and in FSRQs, an extremely efficient
acceleration of relativistic protons within the inner parts of the outflow
is needed, and the jet kinetic power must be orders of magnitude larger 
than the Eddington luminosity \citep{2009ApJ...704...38S}, making this scenario
very unlikely.

\subsection{Interpretation of changes in the SED}
\label{interpretation}

According to the leptonic interpretation of blazar SEDs outlined in 
Sect.~\ref{intro}, the near-IR flux should be dominated by the synchrotron radiation while
the observed $\gamma$-ray flux is a combination of EC and SSC components.
The lack of broad lines in the optical spectrum observed by 
us (Sect.~\ref{sec:keckimgspec}) and 
\cite{2012ApJ...748...49S} suggests that the BLR in GB\,1310$+$487 is weak
or obscured from view. This, however, does not exclude the EC scenario:
photons from the accretion disk or dusty torus might serve as targets
for inverse-Compton scattering. The optical--UV spectrum is flatter than the
near~IR one; however, due to contamination of the host (or intervening) 
galaxy and a nearby star (Sect.~\ref{sec:subsectionNOTimaging}), 
it is not possible to distinguish accretion-disk emission that
might be present in this wavelength range. It may also be that 
the accretion disk avoids detection because it emits at shorter wavelengths.
This would be the case if the central black hole mass is smaller than the one
typically found in blazars, since the accretion-disk temperature decreases
with increasing $M_\bullet$ \citep{1973A&A....24..337S}.
Far-IR data available only during the high-IR state (Flare\,2) are not sufficient 
to estimate the possible contribution from a dusty torus. Thus, accretion-disk and dusty
torus luminosities remain as free parameters in this discussion.

The EC component peaks at higher energies than the SSC component for
the following reasons. First, the typical energy of seed photons for the 
EC process (IR, optical, or UV corresponding to dusty torus, BLR, or accretion disk 
as the dominating source of external radiation) should
be higher than the typical energy of synchrotron photons. The synchrotron
emission peak is located in the far-IR, as suggested by the observed
steep near-IR spectrum. Second, due to relativistic aberration, most
external photons illuminate the synchrotron-emitting plasma blob head on. 
Therefore, the external photons are additionally blueshifted in the reference 
frame of the plasma blob. A change in relative strength of the EC and SSC humps
may explain the observed $\gamma$-ray spectrum evolution, with the harder
spectrum corresponding to greater contribution of the EC component to the total
GeV flux. Large-amplitude GeV variability makes the corresponding
spectral changes apparent even with the limited photon statistics of the
observations.

Another way to explain the observed changes in the $\gamma$-ray spectrum is a
varying contribution from multiple EC components (e.g., EC on 
accretion disk and dusty torus photons). 
This scenario, however, gives no predictions about the behavior 
of the synchrotron SED component, while the SSC$+$EC explanation is able to
describe qualitatively the observed changes in the low-energy SED hump.

The difference in SED during Flare\,1 and Flare\,2 may be 
qualitatively understood if the two flaring events are triggered by different
physical mechanisms. 
The inverse-Compton hump brightening not associated with brightening 
of the synchrotron emission observed in Flare\,1 may result from an increasing
ambient photon field that might be caused by an increasing accretion rate onto
the central supermassive black hole 
\citep{2011ApJ...736..128P}.\footnote{The increased accretion rate should result in accretion-disk 
brightening. However, we may still not detect the accretion disk for the
same reasons we do not detect it in the quiescent state: it is either too
faint or emits at shorter wavelengths. Finally, we cannot exclude the 
possibility that the single available UV data point in the $M2$ band obtained during
Flare\,1 (Table~\ref{table:swiftlog}, Fig.~\ref{figure:sed}) has no contribution 
from the accretion-disk emission.}
Flux increase in both synchrotron and inverse-Compton SED components 
(as observed in GB\,1310$+$487 during Flare\,2), 
combined with a peak energy increase of the two components, may result from
additional electron acceleration. 
The inverse-Compton peak energy increase
(with respect to pre- and post-flare states) is evident during Flare\,2.
The synchrotron peak energy increase during Flare\,2 is not excluded by the
available data.

This fits the pattern of $\gamma$-ray spectrum changes
discussed in Sect.~\ref{discussiongamma}, if the observed $\gamma$-ray
emission is a combination of EC emission peaking at higher energies and SSC
emission peaking at lower energies. The first flare leads to the increased 
EC flux (relative to the SSC flux), which makes the overall $\gamma$-ray spectrum harder. 
The second flare, probably caused by additional electron acceleration, 
is characterized by the enhanced synchrotron (as observed in
the IR) and the corresponding SSC flux, possibly together with the EC
component. The SSC component brightening makes the overall $\gamma$-ray
spectrum softer compared to Flare\,1. The post-flare high-energy hump
might be a combination of SSC and EC components peaking at lower energies
than during Flare\,2 due to lower acceleration of the underlying electron
population. Alternatively, since during the post-flare state the
Compton dominance is less than an order of magnitude, the inverse-Compton
hump might be dominated by the SSC component;
SSC emission with $q>1$ (as in Flare\,2) might be produced
\citep{2012MNRAS.420...84Z,2012ApJ...761..110Z} 
if the energy density of
emitting particles is larger than that of the magnetic field, i.e.,
jet plasma is not in equipartition and Compton losses dominate 
the particle energy loss budget \citep[e.g.,][]{2013MNRAS.431.1840P}.
A constant supply of energy to the emitting particles is needed for this
condition to be fulfilled for an extended period of time \citep{1994ApJ...426...51R}.
Coordinated brightening of both synchrotron and Compton emission components
during Flare\,2 suggests that the SSC mechanism is the one responsible for
the enhanced Compton emission.

It is possible that the lower-energy Compton emission is another EC
component, not the SSC component. For example, the high-energy EC component
responsible for Flare\,1 might be associated with accretion-disk
photons, while the lower-energy EC component could be associated with
lower-energy photons produced in the dusty torus. 
Since the SSC emission has the same beaming pattern as synchrotron emission
\citep[e.g.,][]{2013ApJ...763..134F}, the SSC Compton dominance is independent of $\delta$.
The observed EC radiation intensity has a stronger dependence on $\delta$ than 
does the synchrotron radiation \citep[e.g.,][]{2001ApJ...561..111G}. An increase
in $\delta$ would enhance both synchrotron and EC radiation, and the EC
emission would be enhanced by a larger factor. This would explain the
large value of $q$ during Flare\,2 without the necessity of resorting to
nonstationary SSC models. The SSC peak in this scenario is assumed to be 
at even lower energies than the low-energy EC peak to account for the
spectral change between Flare\,2 and pre/post-flare states. This
``SSC$+$low-energy-EC$+$high-energy-EC" scenario is more complex than
the ``SSC$+$EC" scenario, so we consider the former to be less likely.
Additional detailed SED modeling is needed to test the above scenarios
numerically.

The intermediate interflare state is probably a
combination of the decaying Flare\,1 and rising Flare\,2. 
This may indicate
that either two independent emission zones are responsible for the two
flares, or that a single emission region is propagating through a medium,
with properties gradually changing from those which caused
Flare\,1 to those corresponding to Flare\,2.

The fact that Flare\,1 has no counterpart in the 15\,GHz radio lightcurve
(if a typical value of a few months for the $\gamma$-ray/radio delay is
assumed; see Sect.~\ref{sec:radioproperties})
suggests that the region responsible for Flare\,1 is located upstream of 
the VLBI core region. This region is probably heavily self-absorbed and does
not contribute to the observed flux density at this frequency. Flare\,2 occurred
during the rise of the major radio flare; this may be an indication that the
flaring region that dominates IR and high-energy emission may be
located close to the 15\,GHz radio core, as suggested for other blazars
\citep[e.g.,][]{2010ApJ...715..362J,2012A&A...537A..70S,2012ApJ...758...72W}.

\section{Conclusions}
\label{conclusions}

We summarize the major conclusions of this study as follows.

   \begin{enumerate}
      \item Identification of the flaring $\gamma$-ray source with the
radio source GB\,1310$+$487 is firmly established through positional coincidence and 
simultaneous multiwavelength observations of the flux variability.
      \item Significant changes in the $\gamma$-ray photon index with flux
were observed, showing the harder-when-brighter trend. It may result from a changing
relative contribution of EC and SSC emission to the total $\gamma$-ray flux
in the {\em Fermi}/LAT band. 
      \item The bright near-IR flare does not correspond to the brightest $\gamma$-ray flare. 
This may be an indication that different mechanisms are driving the two observed flares. 
      \item From the absence of VLBA proper motion and the
$\gamma-\gamma$ opacity argument, we constrain the source Doppler 
factor: $1.5 < \delta < 11$.
      \item No clear association could be established between the
$\gamma$-ray variability and changes in radio flux and parsec-scale structure. 
Simultaneous radio/$\gamma$-ray
observations over a longer time baseline are needed to test 
the possibility that some $\gamma$-ray events are associated 
with radio flares.
      \item In the optical band, the object is a blend of
a $\gamma$-ray/radio-loud narrow-line AGN at $z=0.638$ with an unrelated emission-line galaxy
at $z=0.500$. The AGN is not a member of common types of
$\gamma$-ray flaring AGNs (blazars and NLSy1s).
      \item The AGN radiation is probably amplified by a factor of a few
because of gravitational lensing.
   \end{enumerate}

\begin{acknowledgements}

We thank Sara Cutini, Marco Ajello, Denis Bastieri, Boris Komberg, 
Seth Digel, Luca Latronico and the anonymous referee for discussions and comments that helped 
improve this paper.

The \textit{Fermi}/LAT Collaboration acknowledges generous ongoing support
from a number of agencies and institutes that have supported both the
development and the operation of the LAT as well as scientific data analysis.
These include the National Aeronautics and Space Administration (NASA) and the
Department of Energy in the United States, the Commissariat \`a l'Energie Atomique
and the Centre National de la Recherche Scientifique / Institut National de Physique
Nucl\'eaire et de Physique des Particules in France, the Agenzia Spaziale Italiana
and the Istituto Nazionale di Fisica Nucleare in Italy, the Ministry of Education,
Culture, Sports, Science and Technology (MEXT), High Energy Accelerator Research
Organization (KEK) and Japan Aerospace Exploration Agency (JAXA) in Japan, and
the K.~A.~Wallenberg Foundation, and the Swedish Research Council as well as the
Swedish National Space Board in Sweden.
Additional support for science analysis during the operations phase is gratefully
acknowledged from the Istituto Nazionale di Astrofisica in Italy and the
Centre National d'\'Etudes Spatiales in France.

We acknowledge the use of public data from the {\it Swift} data archive at the High Energy
Astrophysics Science Archive Research Center (HEASARC), provided by NASA's
Goddard Space Flight Center.
Based in part on observations with the 100\,m telescope of the MPIfR (Max-Planck-Institut
f\"ur Radioastronomie) and the IRAM 30\,m telescope. IRAM is            
supported by INSU/CNRS~(France), MPG~(Germany) and IGN~(Spain).
The OVRO 40\,m monitoring program is supported in part by NASA grants
NNX08AW31G and NNG06GG1G, and by NSF grant AST-0808050.
This research has made use of data from the MOJAVE database that is 
maintained by the MOJAVE team \citep{2009AJ....137.3718L}.
The data presented herein were obtained in part with ALFOSC, which is
provided by the Instituto de Astrofisica de Andalucia (IAA) under a joint
agreement with the University of Copenhagen and NOTSA.
The MOJAVE project is supported under NASA-Fermi grant NNX08AV67G.

Some of the data presented herein were obtained at
the W. M. Keck Observatory, which is operated as a scientific partnership
among the California Institute of Technology, the University of California,
and NASA. The Observatory was
made possible by the generous financial support of the W. M. Keck Foundation.
We thank O.~Fox, P.~Kelly, I.~Shivvers, and W.~Zheng for
assistance with some of the Keck observations.
The near-IR observations were carried out with the 2.1\,m telescope of
the Guillermo Haro Observatory, INAOE, Mexico.

F.K.S. and K.V.S. were partly supported for this research
through a stipend from the International Max Planck Research School (IMPRS)
for Astronomy and Astrophysics at the Universities of Bonn and Cologne.
I.N. and R.S. are members of the International Max Planck Research School 
(IMPRS) for Astronomy and Astrophysics at the Universities of Bonn and
Cologne.
F.K.S. acknowledges support by the NASA Fermi Guest Investigator program,
grant NXX12A075G.
K.V.S., Y.A.K., and Y.Y.K. were supported in part by the Russian Foundation for Basic
Research (Projects 11-02-00368 and 13-02-12103), the basic research program ``Active
processes in galactic and extragalactic objects'' of the Physical Sciences Division of
the Russian Academy of Sciences, and the Ministry of Education and Science
of the Russian Federation (agreement No.~8405). Y.Y.K. was also supported by the
Dynasty Foundation. RATAN-600 operations were carried out with the financial support of the
Ministry of Education and Science of the Russian Federation (contract
14.518.11.7054). K.V.S. was supported by the Science Education Complex of
the Lebedev Physical Inst. (UNK-FIAN).
A.B.P. was supported by the ``Non-stationary
processes in the Universe'' Program of the Presidium of the Russian Academy
of Sciences.
A.V.F. and S.B.C. are grateful for the support of NASA/{\it Fermi} grant
NNX12AF12GA, NSF grant AST-1211916, the Christopher
R. Redlich Fund, and Gary and Cynthia Bengier.

This research has made use of the NASA/IPAC Extragalactic Database (NED)
which is operated by the Jet Propulsion Laboratory, California Institute of
Technology, under contract with NASA. We also used NASA's Astrophysics Data
System.
K.V.S. thanks Maria Mogilen for her help in preparing this manuscript.

\end{acknowledgements}

\bibliographystyle{aa}
\bibliography{1310}

\end{document}